\documentclass[journal,draftcls,onecolumn,12pt,twoside]{IEEEtranTCOM}
\normalsize

\ifCLASSINFOpdf
\else
\fi

\usepackage{amssymb}
\usepackage{graphicx, tabularx}
\usepackage{amsmath}
\usepackage{algorithm}
\usepackage[noend]{algpseudocode}
\usepackage{setspace, arydshln}
\usepackage{graphicx}
\usepackage{amssymb}
\usepackage{multirow}
\usepackage{slashbox}
\usepackage{picinpar}
\usepackage{booktabs}
\usepackage{flushend}
\usepackage{subfigure}
\usepackage{color}
\usepackage{bm}
\usepackage{multirow}
\usepackage{diagbox}

\begin{document}
\title{Deep Learning Network Based Spectrum Sensing Methods for OFDM Systems}

\author{Qingqing~Cheng,~\IEEEmembership{Student Member,~IEEE,}
        Zhenguo~Shi,\\
        Diep N.~Nguyen,~\IEEEmembership{Member,~IEEE,}
        and~Eryk~Dutkiewicz,~\IEEEmembership{Senior Member,~IEEE}% <-this % stops a space
\IEEEcompsocitemizethanks{\IEEEcompsocthanksitem Q. Cheng, Z. Shi, D. Nguyen, and E. Dutkiewicz are with the School of Electrical and Data Engineering, University of Technology Sydney, Australia.\protect\\
E-mail: \{qingqing.cheng, zhenguo.shi, diep.nguyen, eryk.dutkiewicz\}@uts.edu.au}
%
%
%\thanks{M. Shell is with the Department
%of Electrical and Computer Engineering, Georgia Institute of Technology, Atlanta,
}

\maketitle
\begin{abstract}
%\boldmath
Spectrum sensing plays a critical role in dynamic spectrum sharing, a promising technology to address the radio spectrum shortage. In particular, sensing of orthogonal frequency division multiplexing (OFDM) signals, a widely accepted multi-carrier transmission paradigm, has received paramount interest. Despite various efforts, most conventional OFDM sensing methods suffer from noise uncertainty, timing delay and carrier frequency offset (CFO) that significantly degrade the sensing accuracy. To address these challenges, this work develops two novel OFDM sensing frameworks drawing support from deep learning networks. Specifically, we first propose a stacked autoencoder based spectrum sensing method (SAE-SS), in which a stacked autoencoder network is designed to extract the hidden features of OFDM signals. Using these features to classify the OFDM user's activities, SAE-SS is much more robust to noise uncertainty, timing delay, and CFO than the conventional OFDM sensing methods. Moreover, SAE-SS does not require any prior information of signals (e.g., signal structure, pilot tones, cyclic prefix) which are essential for the conventional feature-based OFDM sensing methods. To further improve the sensing accuracy of SAE-SS, especially under low SNR conditions, we propose a stacked autoencoder based spectrum sensing method using time-frequency domain signals (SAE-TF). SAE-TF achieves higher sensing accuracy than SAE-SS at the cost of higher computational complexity. Extensive simulation results show that both SAE-SS and SAE-TF can achieve significantly higher sensing accuracy, compared with state of the art approaches that suffer from noise uncertainty, timing delay and CFO.
\end{abstract}


% \fi
\IEEEpeerreviewmaketitle

\section{Introduction}
\label{Introduction}
\IEEEPARstart{D}{ynamic} spectrum access (DSA) or spectrum sharing has been widely considered as a promising solution to the radio spectrum shortage \cite{4796930}. Standard bodies like the Federal Communications Commission (FCC) and the European Telecommunications Standardization Institute (ETSI) have been proposing spectrum management frameworks (e.g., Spectrum Access System (SAS) by FCC and Licensed Shared Access by ETSI) that adopt spectrum sharing as a core feature \cite{SAS_LSA}. Under DSA, licensed but underutilized spectrum bands of primary/incumbent users (IUs) are open for secondary users (SUs) with different access priority levels. To avoid harmful interference to IUs as well as to comply with the granted priority right, SUs are required to detect the activity of IUs (e.g., absence or presence). Reliable spectrum sensing allows SUs to occupy or evacuate the spectrum bands, depending on the activity of IUs and other prioritized users{\footnote{Without loss of generality, in this work, we refer to all higher prioritized users (including IUs) as IUs and lower prioritized users as SUs.} \cite{6926773}.
As a widely accepted multi-carrier transmission paradigm, sensing of orthogonal frequency division multiplexing (OFDM) signals has received paramount interest \cite{7577882}.

%In the context of sensing techniques, the spectrum sensing of orthogonal frequency division multiplexing (OFDM) signals has been regarded as a promising candidate and attracts great attention \cite{1542630}.

%The existing signal sensing methods in the literature can be categorized into: non-cooperative and cooperative sensing \cite{8288605}. For the former, the sensing decision (on IU activity) is made solely by a single SU. For the latter, each SU makes a local decision and all those decisions are reported to a fusion center to achieve a final decision based on some fusion rules \cite{8093693}. In this work, we focus on the non-cooperative OFDM sensing methods.

The energy detection (ED) is one of the simplest and most popular sensing methods, which detects the IU's activity based on the energy of the received signals \cite{6914542}. To leverage the special features of OFDM signals, e.g., pilot tones (PT) \cite{NPSSOSESDBS}, cyclic prefix (CP) \cite{osssoskunv,7748543} and covariance matrix (CM) features \cite{7265098}, one can use the feature-based detection approach. For instance, \cite{osssoskunv} presents methods to detect the IU's signals by exploiting the autocorrelation of CP. In \cite{7265098}, authors propose a CM-based method to determine the IU's activity states by leveraging the features of the covariance matrix of the discrete Fourier transform of the received signals. However, the sensing accuracies of these methods are heavily dependent on the noise uncertainty, carrier frequency offset (CFO) or synchronization errors/timing delay \cite{7883952,7147789}. Moreover, those methods require full or partial prior knowledge of IU's signals (e.g., the CP or PT structure of IU signals) and/or noise power that are unavailable in some practical applications (e.g., when IUs are military applications) \cite{7889011}. Instead of requiring features as a priori knowledge, in this work, we employ the latest advances in machine learning (ML) \cite{7307098} to learn them. More importantly, our methods can also learn/capture the hidden features of OFDM signals to improve the sensing accuracy.

ML has recently found its applications in various areas such as object detection \cite{7004814}, speech recognition \cite{6423821,8085174}, channel estimation \cite{8052521}, and pattern recognition \cite{8360460,6399478}. We observe that spectrum/signal sensing resembles a pattern recognition problem, as illustrated in Fig. \ref{spectrum sensing vs pattern recognition}. Specifically, pattern recognition involves the steps of the feature extraction and the classification \cite{6399478}. Analogously, typical OFDM sensing methods consist of two steps: first, calculate the test statistic of the received signals; second, compare the test statistic with the corresponding thresholds to detect IU's activity. We then can map these first and second stages in OFDM sensing to the feature extraction and the classification in pattern recognition, respectively.
\begin{figure}[t]
\setlength{\abovecaptionskip}{0.cm}
\setlength{\belowcaptionskip}{-0.cm}
\centering
\includegraphics[width=10 cm]{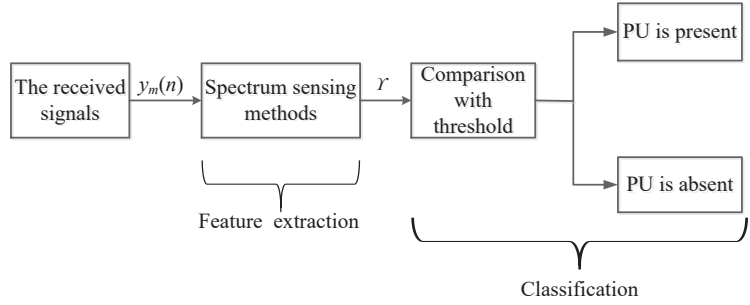}
\DeclareGraphicsExtensions.
\caption{Relationship between spectrum sensing and pattern recognition}
\vspace{-1cm}
\label{spectrum sensing vs pattern recognition}
\end{figure}

Most ML-based spectrum sensing works have been focusing on cooperative spectrum sensing (CSS) \cite{6336689} that utilizes ML to fuse individual sensing results from multiple SUs for decision process. The authors of \cite{5751185} develop a linear fusion rule for CSS, which utilizes the Fisher linear discriminant analysis to obtain linear coefficients. {{The authors in \cite{6635250} propose several cooperative sensing algorithms based on support vector machine (SVM), weighted K-nearest-neighbor (KNN), K-means clustering, and Gaussian mixture model (GMM). They use the energy of received signals as feature vectors. The approach in \cite{7841503}, utilizing the fuzzy SVM and nonparallel hyperplane SVM, is also claimed to be robust to the noise uncertainty. Although those cooperative sensing methods can improve the sensing performance, their complexities are significantly high. To solve that problem, in \cite{DBLP:journals/corr/LeeKCS17a}, a convolutional neural network (CNN) based cooperative sensing method is proposed, which improves the sensing performance with low complexity. In \cite{7564840}, the K-means clustering and SVM techniques are used but taking a low-dimensional probability vector as the feature vector, resulting in a smaller training duration and a shorter classification time. Besides, a ML-based mobile CSS framework for large-scale heterogeneous cognitive radio networks is proposed in \cite{8466022}, drawing support from the recent advances in Bayesian machine learning.

Unlike the above cooperative spectrum sensing methods, our work provides non-cooperative spectrum sensing solutions using deep learning in which very little has been investigated so far. The only and most relevant to ours is \cite{8292449} that proposes a sensing method based on Artificial Neural Network (ANN), utilizing the energy and Likelihood Ratio Test statistic as input features. In \cite{5601105}, the authors also apply ANN for sensing purpose with the energy and cyclostationary features as the input features. Using the same features as \cite{5601105}, the authors in \cite{8302117} rely on the CNN architecture instead.}}

Although these methods help improve the sensing performance in non-cooperative scenarios, they all need to rely on explicitly extracted features. Consequently, the accuracy of input features that are explicitly extracted from the received signals would directly influence the sensing results. In other words, their performance are strongly dependent on the external feature extraction algorithms. Moreover, extracting specific or known features from the original received signal can only obtain partial information. This is because the explicit feature extraction process inevitably loses information of implicitly hidden but helpful features, degrading the sensing performance.

In this work, we leverage deep learning networks to address all the above limitations. Instead of manually describing the event of interest with explicit features, deep learning (DL) \cite{SCHMIDHUBER201585} relies on multiple layers of nonlinear processing units (so called a deep architecture) to extract both known as well as possibly hidden features of the input signals.

There are various deep learning network architectures, e.g., recurrent neural network (RNN), convolutional neural network (CNN), stacked autoencoder, etc \cite{8382166}. Among these, in this work we adopt the stacked autoencoder (SAE) \cite{8367975} for the following reasons. First, RNN is a generative model in which the ``output" is taken to be the predicted input data in the future. RNN has been widely used in prediction-related work, e.g., in modeling the speech data. However, it is extremely difficult to train RNN properly due to the well known ``vanishing gradient" problem \cite{Martens:2011:LRN:3104482.3104612}. CNN is a partial connection model containing {{{convolutional layers and pooling layers}}}. CNN is particularly helpful for applications with geometry features, e.g., computer vision \cite{SCHMIDHUBER201585}. However, the feature extraction process in CNN inevitably loses information due to its partial connection model. By contrast, SAE is a fully connected network consisting of {{{encoders and decoders}}} \cite{8367975}. The encoder is able to effectively (but implicitly) extract and learn the essential information/feature that captures the main variations of its input data. Moreover, it also can detect and remove the input redundancies while preserving only the essential aspects of the data. Utilizing those extracted features, the decoder is able to effectively reconstruct the actual input data so that the reconstructions are as similar as possible to the actual input data \cite{NIPS2006_3048}. Second, compared with other deep learning networks (e.g., CNN and RNN), the SAE architecture is conceptually simple and easy to be trained \cite{7163353}. It can be trained through a greedy layer-wise unsupervised pre-training followed by supervised fine-tuning process \cite{10.1007/978-3-642-21735-7_7}. In the unsupervised pre-training phase, each layer is trained through the encoding-decoding process to extract the essential information of the data. In the supervised fine-tuning, the back-propagation method is used to adjust and optimize parameters of the whole network, improving the accuracy of classifying different extracted information \cite{Erhan:2010:WUP:1756006.1756025}. Intensive simulations show that the two proposed SAE-based sensing methods are robust to noise uncertainty, timing delay and carrier frequency offset (CFO). The major contributions of this work are summarized below:
\begin{itemize}\vspace{-0.5cm}
\item We propose a stacked autoencoder based spectrum sensing method (SAE-SS) to extract hidden features of the original received signals and detect the IU’s activities based on the extracted features. Compared with existing sensing methods (i.e., both conventional and learning-based methods), SAE-SS is more robust to timing delay, noise uncertainty and CFO.

\item Unlike most existing ML-based methods, SAE-SS does not require any prior knowledge of the IU's signal structure nor any external feature extraction algorithm. Instead, it automatically extracts relevant features of the IU's presence/absence from the received signal.

\item To further improve the sensing accuracy of SAE-SS, especially under low SNR conditions, we incorporate the frequency domain information into the SAE-SS. By relying on both the received signal and its presentation in the frequency domain using the fast Fourier transform (FFT), the new method, called SAE-TF, achieves higher sensing accuracy at the cost of higher computational complexity (compared with SAE-SS).

\item We provide comprehensive analysis on the computational complexity, training time, and the feasibility of SAE-SS and SAE-TF, in comparison with state of the art OFDM spectrum sensing methods. We also analyze the performance of proposed methods under different conditions (e.g., CP length, hidden units, OFDM blocks, and additional features).

\item To evaluate the performance of the proposed methods, we implement SAE-SS and SAE-TF using the TensorFlow $1.3$ software with Python language. The simulation results show that our proposed methods achieve significantly higher sensing accuracy than the state of the art OFDM sensing methods, even in the low SNR regime.
\end{itemize}

The remainder of the paper is organized as follows. Section \ref{System Model and Problem Formulation} describes the system model and the problem formulation. Section \ref{Proposed Stacked Autoencoder Based Spectrum Sensing Method} and Section \ref{Proposed Joint Stacked Autoencoder Based Spectrum Sensing Method} present the design details of SAE-SS and SAE-TF, respectively. Complexity analysis and other discussion are described in Section \ref{Complexity Analysis and Other Discussions}. Simulation results are shown in Section \ref{Simulation Results}, followed by conclusions in Section \ref{Conclusion}.
\section{System Model and Problem Formulation}\label{System Model and Problem Formulation}
\subsection{System Model}
Let $y_m(n)$ denote the received OFDM signals at the SU receiver, where $m\in[0,\dots,M]$ is the total number of received OFDM blocks and $n\in[0,\dots,N_c+N_d-1]$. $N_c$ denotes the length of the cyclic prefix and $N_d$ denotes the data block size. $y_m(n)$ can be written as:
\begin{eqnarray}\label{ymn}
&&H_0:y_m(n)\!=\!w_m(n),
\nonumber\\
&&H_1:y_m(n)\!=\!e^{\!-j\frac{2\pi f_q(n\!-\!\delta)}{N_d}}\!\!\sum_{l\!=\!0}^{L_p\!-\!1}\!\!h_ls_m\!(n\!-\!\delta\!-\!l)\!\!+\!\!w_m\!(n),
\end{eqnarray}
where $H_0$ and $H_1$ represent the hypotheses of absence and presence of IU, respectively. $w_m(n)$ is the complex additive white Gaussian noise (AWGN) with zero-mean and variance $\sigma_w^2$ (i.e., $w_m(n)\sim CN(0,\sigma_w^2)$). $s_m(n)$ denotes the transmitted OFDM signals by the IU transmitter. $f_q$ is the normalized carrier frequency offset. $\delta$ denotes the timing delay. $h_l$ represents the channel gain of the $l$th channel and we assume its value does not change during the sensing process. $L_p$ presents the total number of multi-path components between the IU and the SU. Without loss of generality, $w_m(n)$, $s_m(n)$ and $h_l$ are assumed to be mutually independent. According to the central limit theorem, when the length of the received signals is sufficiently large, $s_m(n)$ approximately obeys the complex Gaussian distribution, and its mean and variance are zero and $\sigma_s^2$, respectively. In that case, SNR of $y_m(n)$ under $H_1$ can be represented as $\text{SNR}=\sigma_s^2\sum_{l=0}^{L_{p}-1}{|h_l|^2}/\sigma_w^2$.

\subsection{Problem Formulation}
For typical OFDM sensing methods, e.g., the basic ED, its test statistic $\it{\Upsilon}_{\text{ED}}$ is:
\begin{eqnarray}\label{ten}
\it{\Upsilon}_{\text{ED}}=\frac{1}{M(N_c+N_d)}\sum_{m=0}^{M-1}\sum_{n=0}^{N_c+N_d-1}{|y_m(n)|^2}.
\end{eqnarray}

The probability of false alarm (PFA) and probability of missed detection (PM) are
\begin{eqnarray}\label{pen}
\text{PFA}_{\text{ED}}=\text{Pr}(\it{\Upsilon}_{\text{ED}}>\lambda|H_0),
\quad\text{PM}_{\text{ED}}=\text{1}-\text{Pr}(\it{\Upsilon}_{\text{ED}}>\lambda|H_1),
\end{eqnarray}
where $\text{Pr}(.)$ denotes the probability distribution function. $\lambda$ is the sensing threshold based on the estimated noise variance ($\widetilde {\sigma}_w^2$). According to above equations, ED only utilizes the received signals to detect the presence of IU without requiring any prior knowledge of IU signals. However, it is vulnerable to noise uncertainty. Moreover, even a small estimated error in the noise power would significantly deteriorate its sensing accuracy, especially in the low SNR regime \cite{7756405}.

For feature-based detections, e.g., CP-based \cite{osssoskunv} and CM-based \cite{7265098}, their test statistics are
\begin{eqnarray}\label{tcp}
&&{{\it{\Upsilon}_{\text{CP}}=\max\limits_{\delta}\text{log}\Big(\frac{\text{Pr}(\mathbf{R}|H_1,\sigma_w^2,\sigma_s^2)}{\text{Pr}(\mathbf{R}|H_0,\sigma_w^2)}\Big) }},
\nonumber\\
&&{\it{\Upsilon}}_{\text{CM}}\!=\!\frac{||\hat{\mathbf{R}}\odot(\mathbf{1}_{N_d}-\mathbf{I}_{N_d})||_{L_1}}{\sqrt{N_d^2-N_d}},
\end{eqnarray}
where $\mathbf{R}$ is the correlation vector of the received signals. $\hat{\mathbf{R}}$ is the covariance matrix of received signals after discarding the CP in frequency domain.
%where $\mu_{1,i}$ is the mean of $R_i|H_1$. $\sigma_{0,i}^2$ is the variance of $R_i|H_0$. , and $(.)^*$ represents the complex conjugate operation.

Although CP-based and CM-based detections are robust to noise uncertainty, they require some prior knowledge of the IU's signals, such as the IU's transmitting power or the CP structure. However, this information is usually not available. Moreover, as aforementioned, the feature-based detections are sensitive to timing delay and CFO, and even a small mismatch between the received and transmitted signals in the time domain or frequency domain would lead to significant degradation of sensing accuracy.

%To overcome the drawbacks associated with traditional OFDM sensing methods, we propose a novel spectrum sensing methods based on Stacked SAE network.

\section{Stacked Autoencoder Based Spectrum Sensing}
\label{Proposed Stacked Autoencoder Based Spectrum Sensing Method}
In this section, we propose SAE-SS that is robust to timing delay, noise uncertainty and CFO. The architecture of the OFDM system with SAE-SS is shown in Fig. \ref{architecture_of_the_OFDM system_with_stacked_autoencoder_based_spectrum_sensing}. The hidden features of received OFDM signals are extracted from the SAE network during the offline training stage. Those extracted features are then used to sense IU's activity during the online sensing stage. The key steps of SAE-SS can be summarized as: pretraining, fine-tuning of SAE, and classification (e.g., online spectrum sensing stage).
\begin{figure}[tp]
\setlength{\abovecaptionskip}{0.cm}
\setlength{\belowcaptionskip}{-0.cm}
\centering
\includegraphics[width=10 cm]{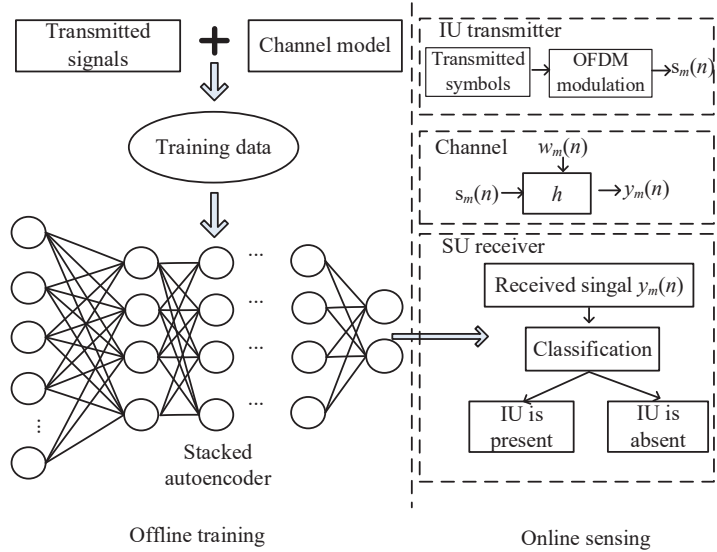}
\DeclareGraphicsExtensions.
\caption{Architecture of the OFDM system with SAE-SS}
\vspace{-1cm}
\label{architecture_of_the_OFDM system_with_stacked_autoencoder_based_spectrum_sensing}
\end{figure}

The spectrum sensing technique (both conventional and learning-based) can be described as a process of the feature extraction followed by the feature classification. The feature extraction aims to extract the essential features for sensing, which can be expressed as
\begin{eqnarray}\label{feature extraction of SS}
{\bf{\Upsilon}}=F(\mathbf{y}),
\vspace{-1cm}
\end{eqnarray}
where ${\bf{\Upsilon}}$ denotes the extracted feature vector, ${\bf{\Upsilon}} \in {{\mathbb{R}}}^{d_{\it{\Upsilon}}}$, $d_{\it{\Upsilon}}$ is the dimension of the feature vector. $\mathbf{y}$ stands for the received signal vector, $\mathbf{y}:=\{y_0(0),y_0(1),\dots,y_m(n),\dots,y_M(N_c+N_d-1) \}$, $\mathbf{y} \in \mathbb{R}^{d_{y}}$, $d_y=M(N_c+N_d)$. $F(.)$ is the function of feature extraction. Then the extracted features will be classified into different classes for sensing the IU's presence. Note that for different sensing methods, the meanings of ${\bf{\Upsilon}}$ and $F(.)$ are different.

For the conventional sensing methods, $\bf{{\Upsilon}}$ in equation (\ref{feature extraction of SS}) denotes the test statistic of specific sensing methods, which is usually one dimensional feature, denoted as ${{\it{\Upsilon}}}$. $F(.)$ stands for the function that calculates the test statistic, which is determined by the specific sensing methods with the certain models. Upon achieving ${\it{\Upsilon}}$ calculated by $F(.)$, SUs can detect the IU's presence by just comparing ${\it{\Upsilon}}$ with a threshold. Take the energy detection as an instance, ${\it{\Upsilon}}_{ED}$ is the test statistic that is one-dimensional feature, and $F(\bf{.})$ is a function to calculate ${\it{\Upsilon}}_{ED}$, ${\it{\Upsilon}}_{ED}=F(\mathbf{y})=\frac{1}{d_y}\mathbf{y}\mathbf{y}^{T}$. If ${\it{\Upsilon}}_{ED}$ is smaller than the threshold, then IU is absent, otherwise, IU is present.

For the learning-based sensing methods, ${\bf{{\Upsilon}}}$ in equation (\ref{feature extraction of SS}) is the feature vector extracted by network, and $F(\bf{.})$ is the network structure. Take the proposed SAE-SS as an example, ${\bf{{\Upsilon}}}_{SAE}$ is the feature vector extracted by SAE. $F_{SAE}(\bf{.})$ is the SAE network structure which is parameterized by $\{\bf{W, b}\}$, $\bf{W}$ stands for the weight matrix and $\bf{b}$ is the bias matrix. Generally, ${\bf{{\Upsilon}}}_{SAE}$ contains multiple features that involve IU's activity states (absence and presence). Thus a classifier is used to classify these features into separated classes for detecting the IU's presence. The basic idea is to utilize the SAE as a feature extractor to automatically extract the essential hidden features from the original input signal, without any external feature extraction algorithms. To that end, we need to train the SAE network for feature extraction.

The key idea of SAE-SS is to train the SAE network for extracting the hidden features from the original received signal. This training process is conducted during the offline stage (i.e., before the spectrum sensing phase), which includes the pre-training and fine-tuning. The trained SAE is then directly used to detect IU's activity states during the online sensing phase.
\subsection{Pretraining of Stacked Autoencoder }
\begin{figure*}[tb]
\setlength{\abovecaptionskip}{0.cm}
\setlength{\belowcaptionskip}{-0.cm}
\centering
\subfigure[Structure of AE]{
\label{fig1_a}
\includegraphics[width=0.28\textwidth]{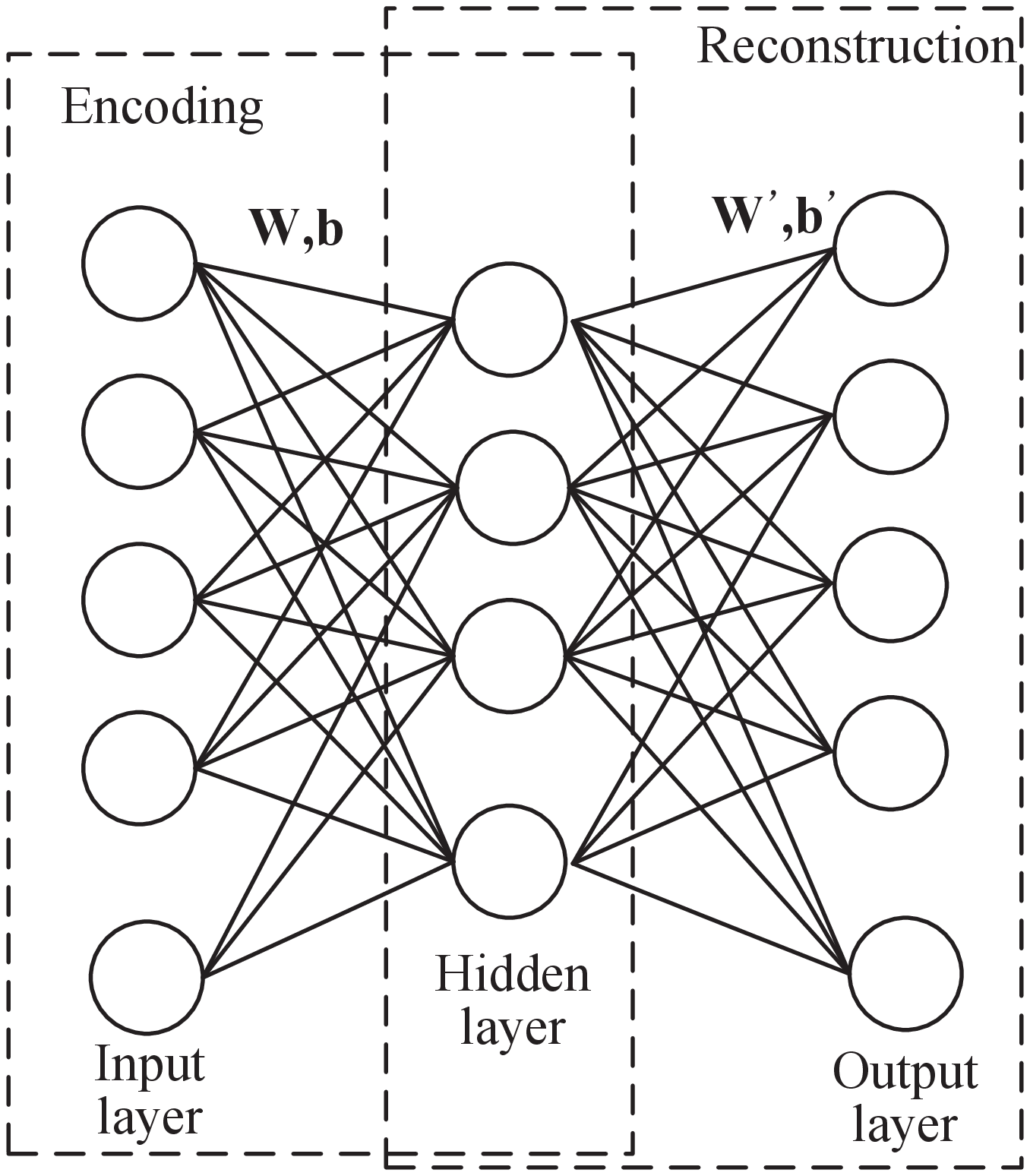}}
\hspace{1in}
\subfigure[Structure of SAE with logistic regression classifier]{
\label{fig2_a}
\includegraphics[width=0.51\textwidth]{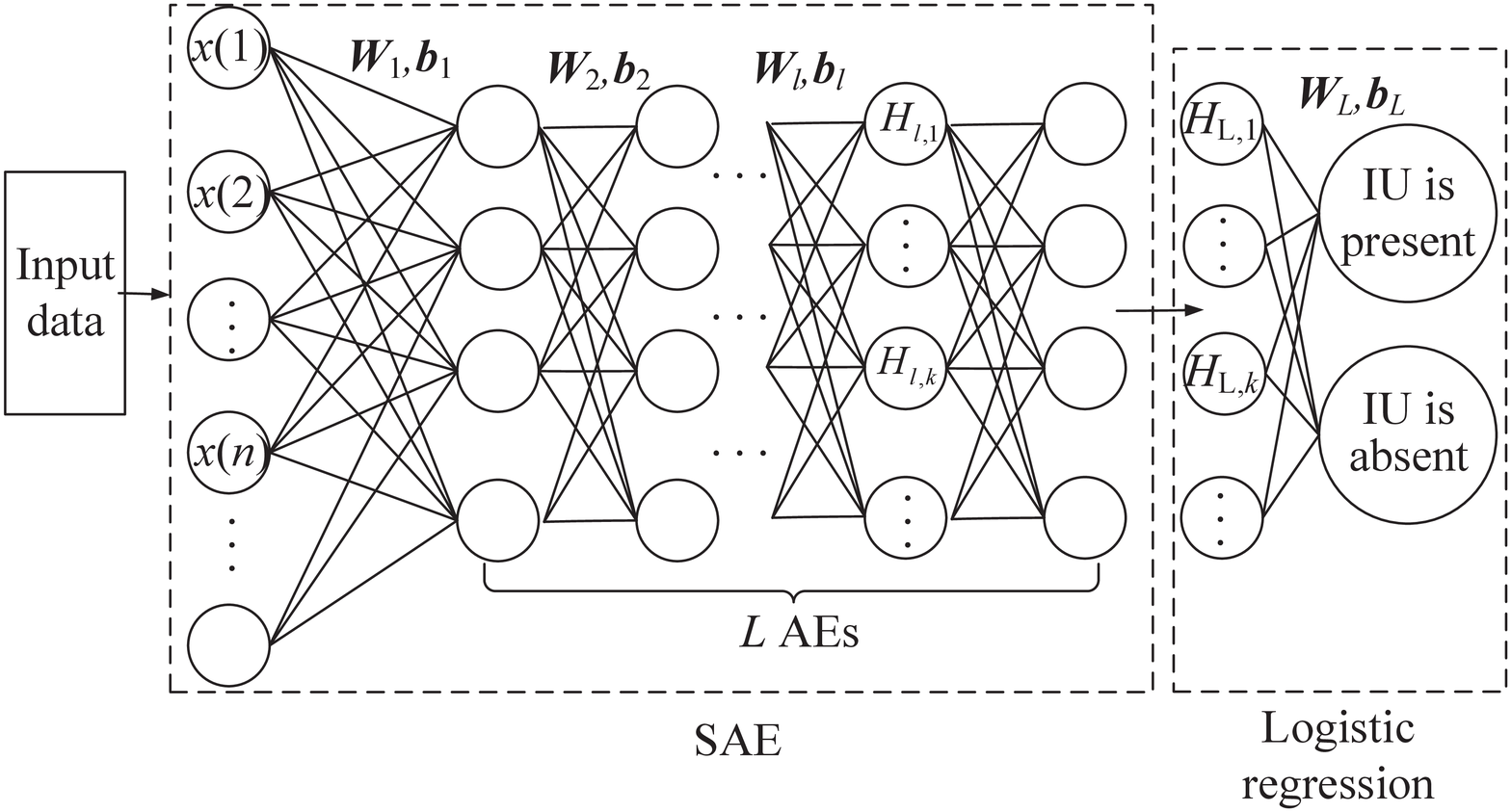}}
\caption{The structure of spectrum sensing based on traditional neural networks}
\vspace{-1.235cm}
\label{fig2}
\end{figure*}
The pretraining SAE aims to learn the hidden features of received signals through two stages. In the first stage, the SAE network is divided into independent Autoencoders (AEs) and these AEs are individually trained, one by one. Each AE is a three-layer network including input layer, hidden layer and reconstruction layer, as illustrated in Fig. \ref{fig1_a}. In the second stage, the input and hidden layers of the trained AEs are stacked together, layer by layer, as shown in Fig. \ref{fig2_a}.

The training of each AE can be described as an encoding-decoding process. The input data is first mapped into a hidden presentation via an encoding process. In addition to known features (e.g., energy and test statistics features), this presentation also contains other useful yet unnamed/unknown information/features for the sensing purpose \cite{6844831}. These features are referred to as the hidden/unknown features. The resulting latent presentation is then mapped back to a ``reconstructed" vector through a decoding process. The encoding-decoding process aims to minimize the average reconstruction error.

Note that AE is only suitable for non-complex numbers, thus we partition the input signals into real and imaginary parts, respectively. The received signal in equation (\ref{ymn}) is rewritten as:
\begin{eqnarray}\label{y_i}
\mathbf{y}_i\!\!\!&:=\!\!&\!\!\!\{\Re({y}_i(0)),\Im({y}_i(0)),\Re({y}_i(1)),\Im({y}_i(1)),\!\dots\!,
\Re({y}_i(N_c+N_d-1)),\Im({y}_i(N_c\!+\!N_d\!-\!1))\},
\end{eqnarray}
where $\Re(.)$ and $\Im(.)$ mean the real part and the imaginary part, respectively. Thus the input vector of SAE-SS network $\mathbf{x}$ can be expressed as
\begin{eqnarray}\label{input_x}
\mathbf{x}:=\{\mathbf{y}_0, \mathbf{y}_1, \dots, \mathbf{y}_{M-1}\}^T,
\end{eqnarray}
where the length of $\mathbf{x}$ is $N_{input}=2M(N_c+N_d)$. Then $\mathbf{x}$ is used to train AEs. Take the $l$th AE as an example, let $H_{l,p}^{in}$, $H_{l,k}$ and $H_{l,q}^{out}$ denote the $p$th input unit, the $k$th hidden unit and the $q$th output unit, respectively. Then $H_{l,k}$ can be obtained with $H_{l,p}^{in}$ \cite{MAL-006}, which is shown as below:
\begin{eqnarray}\label{a5}
H_{l,k}=f(\sum_{p=1}^{P_l}W_{l,p,k}H_{l,p}^{in}+b_{l,k}),
\end{eqnarray}
where $P_l$ is the number of input units in the $l$th AE. $f(.)$ is activation function. $W_{l,p,k}$ denotes the weight between the $p$th input unit and the $k$th hidden unit of the $l$th AE. $b_{l,k}$ is the bias of the $k$th input unit of the $l$th AE. Then $H_{l,q}^{out}$ can be achieved with $H_{l,p}^{in}$ and $H_{l,k}$, by
\begin{eqnarray}\label{a6}
&&H_{l,q}^{out}=f(\sum_{k=1}^{K_{l}}W_{l,k,q}^{'}H_{l,k}+b_{l,q}^{'})
=f\Bigg(\sum_{k=1}^{K_{l}}W_{l,k,q}^{'}f\bigg(\sum_{p=1}^{P_l}W_{l,p,k}H_{l,p}^{in}+b_{l,k}\bigg)\!+\!b_{l,q}^{'}\Bigg),
\end{eqnarray}
where $K_l$ is the number of units in the hidden layer of the $l$th AE. $W_{l,k,p}^{'}$  denotes the weight between the $k$th hidden unit and the $q$th output unit of the $l$th AE. $W_{l,p,k}=W_{l,k,q}^{'}$. $b_{l,q}^{'}$ is the bias of the $q$th output unit of the $l$th AE.
When $l=1$, then the input vector $\mathbf{H}_{1}^{in}$ is same as $\mathbf{x}$, which is
{\setlength\abovedisplayskip{3pt}
\setlength\belowdisplayskip{3pt}
\begin{eqnarray}\label{input_H1}
\mathbf{H}_{1}^{in}:=\{{H}_{11}^{in}, {H}_{12}^{in}, \dots, {H}_{1P_1}^{in}\}^T,
\end{eqnarray}
}
where $P_{1}$ is the number of input units in the first AE, $P_{1}=N_{input}=2M(N_c+N_d)$.
In this paper, the sigmoid function \cite{7163353} is adopted as the activation function, which is
\begin{eqnarray}\label{a2}
f(H_{l,p}^{in})=\frac{1}{1+e^{-H_{l,p}^{in}}}.
\end{eqnarray}

The aim of training the $l$th AE is to minimize the error between $H_{l,p}^{in}$ and $H_{l,p}^{out}$ by continuously updating the values of $W_{l,p,k},b_{l,k}$ and $b_{l,q}^{'}$. The most typical error function to measure the difference between $H_{l,p}^{in}$ and $H_{l,p}^{out}$ is the mean square error:
\begin{eqnarray}\label{a7}
\chi=\frac{1}{P_l}\sum_{p=1}^{P_l}||H_{l,p}^{in}-H_{l,p}^{out}||^2.
\end{eqnarray}

However, this method is time-consuming for training, thus we select the cross-entropy method \cite{8252873} to speed up the training process, which is
\begin{eqnarray}\label{a8}
\chi=\sum_{p=1}^{P_l}[H_{l,p}^{in}\mathrm{log}(H_{l,p}^{out})+(1-H_{l,p}^{in})\mathrm{log}(1-H_{l,p}^{out})].
\end{eqnarray}

Let $\Omega=\{W_{l,p,k},b_{l,k},b_{l,q}^{'}\}$, then the objective function is
\begin{eqnarray}\label{SAE-OFDM: 10}
\Omega=\arg\min_{\Omega}{\chi}.
\end{eqnarray}

Moreover, we adopt the gradient descent method \cite{doi:10.1137/1.9781611970920.bm} to achieve the optimal $\Omega$, by
\begin{eqnarray}\label{SAE-OFDM: 11}
&&W_{l,p,k}(n+1)\leftarrow W_{l,p,k}(n)-\kappa\frac{\partial \chi(H_{l}^{in},H_{l}^{out})}{\partial W_{l,p,k}},
\nonumber\\
&&b_{l,k}(n+1)\leftarrow b_{l,k}(n)-\kappa\frac{\partial \chi(H_{l}^{in},H_{l}^{out})}{\partial b_{l,k}},
\nonumber\\
&&b_{l,q}^{'}(n+1)\leftarrow b_{l,q}^{'}(n)-\kappa\frac{\partial \chi(H_{l}^{in},H_{l}^{out})}{\partial b_{l,q}^{'}},
\end{eqnarray}
where $W_{l,p,k}(n)$, $b_{l,k}(n)$ and $b_{l,q}^{'}(n)$ denote weight and bias of $n$th training. $\kappa$ is the leaning rate.

Upon training all the AEs based on the above rules, the trained SAE is created by stacking the input and hidden layers of trained AEs together, layer by layer. The output of SAE's pre-training, which is also the hidden units of $L$th AE, can be written as
\begin{eqnarray}\label{input_HL}
{\mathbf{H}}_L:=\{H_{L1}, H_{L2}, \dots, H_{LK_{L}}\}^T,
\end{eqnarray}
where $K_L$ is the number of units in the hidden layer of the $L$th AE. $\mathbf{H}_{L}$ in equation (\ref{input_HL}) contains the hidden features of the received signals and will be used to sense the IU's activity states.
\subsection{Fine-Tuning and Classification}
The pretraining process of SAE can be interpreted as extracting the IU's unsupervised features. Thus, the trained SAE needs to be fine-tuned to leverage the SAE's property for the spectrum sensing. In this paper, we select the logistic regression classifier \cite{6757003} to fine-tune the SAE network, as shown in Fig. \ref{fig2_a}.
\begin{table}\centering
	\label{SAE-SS Steps}
	\vskip 0.3cm
\begin{minipage}[htbp]{3.83 in}
\rule{\linewidth}{0.3mm}\vspace{-0.05in}
{{\bf{Algorithm 1}}: Stacked Autoencoder Based Spectrum Sensing Method (SAE-SS)}\vspace{-0.12in}\\
\rule{\linewidth}{0.2mm}
{ {\small
\begin{tabular}{ll}
   1:&\textbf{begin}\\
   2:&\textbf{Initialize}: the length of input $N_{input}$, number of SAE\\
     & layer $L$, the number of hidden unit in $l$th layer $K_l$, \\
     & pretraining iterations $N_{pr}$, pre-training learning rate $\kappa_p$, \\
     & fine-tuning iterations $N_f$, fine-tuning rate $\kappa_f$, \\
     & number of classes $C$; \\
   3:& Achieve $\mathbf{x}$ based on $\mathbf{y}$\\
   4:& \textbf{For} $1\leq l\leq L$\\
   5:& \quad  Build an AE with $N_v$ units of input layer and $N_h$ unit\\
     & \quad of hidden layer;  \\
   6:& \quad  \textbf{If} $l=1$\\
   7:& \quad \quad $N_v=N_{input}$; $N_h=K_1$; $\mathbf{x}$ is set as the input of AE;\\
   8:& \quad \textbf{else} \\
   9:& \quad \quad $N_v=P_l$; $N_h=K_l$;       \\
   10:& \quad \quad the hidden units of previous layer ${\mathbf{H}}_{l-1}$ is set as the \\
      & \quad \quad input of current layer ${\mathbf{H}}_{l}^{in}$;      \\
   11:& \quad \textbf{end} \\
   12:& \quad Initialize AE, generate $\mathbf{W}_l$, $\mathbf{b}_l=\mathbf{b}_l^{'}=0$\\
   13:& \quad \textbf{For} $1\leq t\leq N_{pr}$\\
   14:& \quad \quad Based on (\ref{a6}) calculate the output of reconstruction:\\
      & \quad \quad$H_{l,q}^{out}=f(\sum_{k=1}^{K_{l}}W_{l,k,q}^{'}H_{l,k}+b_{l,q}^{'})$\\
   15:& \quad \quad Based on (\ref{a8}) calculate the error:\\
      & \quad \quad $\chi\!=\!\sum_{p=1}^{P_l}[H_{l,p}^{in}\!\mathrm{log}(H_{l,p}^{out})\!+\!(1\!-\!H_{l,p}^{in})\!\mathrm{log}(1\!-\!H_{l,p}^{out})]$\\
   16:& \quad \quad Based on~\eqref{SAE-OFDM: 11} update $\mathbf{W}_l$, $\mathbf{b}_l$ and $\mathbf{b}_l^{'}$ with $\kappa_p$\\
   17:& \quad \textbf{end}    \\
   18:& \quad remove the reconstruction layer of AE\\
   19:& \textbf{end} \\
   20:& Initialize logistic regression layer: $K_L$ unit of input layer,\\
      & \quad $C$ unit of output layer\\
   21:& \quad\textbf{For} $1\leq t\leq N_f$\\
   22:& \quad\quad Based on~\eqref{SAE-OFDM: 14} calculate probability of each class\\
   23:& \quad \quad Based on back propagation, update parameters of \\
      & \quad \quad every layer with $\kappa_f$\\
   24:& \quad \textbf{end}\\
   25:&\textbf{end}\\
\end{tabular}}}\\
\rule{\linewidth}{0.3mm}
\end{minipage}\\
\vspace{-1cm}
%\caption{Summary of Key Steps in PSEO}
\end{table}

The logistic regression classifier can be regarded as a neural network with a single layer, and the activation of output layer is the softmax function. The input of the logistic regression classifier is $\mathbf{H}_{L}$, which is the output of the pretraining SAE. $U$, the output of logistic regression classifier, can be seen as a set of conditional probabilities of $\mathbf{H}_{L}$, $\mathbf{W}_R$ and $\mathbf{b}_R$. $\mathbf{W}_R$ and $\mathbf{b}_R$ are weights and biases of the logistic regression layer. Let $\tau=1$ and $\tau=0$ denote the presence and the absence of IU activity states, respectively. According to the logistic regression classifier \cite{6757003}, then the conditional probability of $U$ is:
\begin{eqnarray}\label{SAE-OFDM: 14}
\text{Pr}(U\!=\!\tau|\mathbf{H}_{L},\mathbf{W}_R,\mathbf{b}_R)\!=\!\frac{e^{\mathbf{W}_{R,\tau}\mathbf{H}_{L}+\mathbf{b}_{R,\tau}}}
{\sum_{i=0}^{1}{e^{\mathbf{W}_{R,i}\mathbf{H}_{L}+\mathbf{b}_{R,i}}}}.
\end{eqnarray}

We apply the back-propagation method \cite{7293666} to train the logistic regression classifiers. The whole SAE network is also fine-tuned at the same time. Upon pretraining and fine-tuning the SAE network in the offline phase, SAE-SS can detect the IU's activity states using only the received signals. In comparison with the feature-based OFDM spectrum sensing methods, the inputs of SAE-SS are the originally received signals. SAE-SS does not require any arithmetical operations such as calculating the energy or the correlation values. Consequently, the overhead of SAE-SS during the online sensing stage is reduced significantly. Moreover, the proposed SAE-SS can complete the sensing task without any prior knowledge of the IU's information, which is more suitable for the practical environment. That is particularly relevant to military applications, opening up the military radar bands for secondary use. The procedure of SAE-SS is summarized in the pseudo-code in Algorithm 1.
\section{Stacked Autoencoder Based Spectrum Sensing with Time-Frequency Domain Signals}
\label{Proposed Joint Stacked Autoencoder Based Spectrum Sensing Method}
As aforementioned, the proposed SAE-SS achieves high sensing accuracy by extracting all hidden features of the received OFDM signals without requiring any external feature extraction algorithms. However, its sensing accuracy would degrade under low SNR conditions. To address that problem, we present a Stacked Autoencoder Based Spectrum Sensing Method with time-frequency domain signals (SAE-TF). The input data of SAE-TF involves both time domain and frequency domain signals, which are beneficial for SAE to extract more hidden features for the spectrum sensing purpose. The framework of SAE-TF is shown in Fig. \ref{SAE-TF}.
\begin{figure}[t]
\setlength{\abovecaptionskip}{0.cm}
\setlength{\belowcaptionskip}{-0.cm}
\centering
\includegraphics[width=9.2 cm]{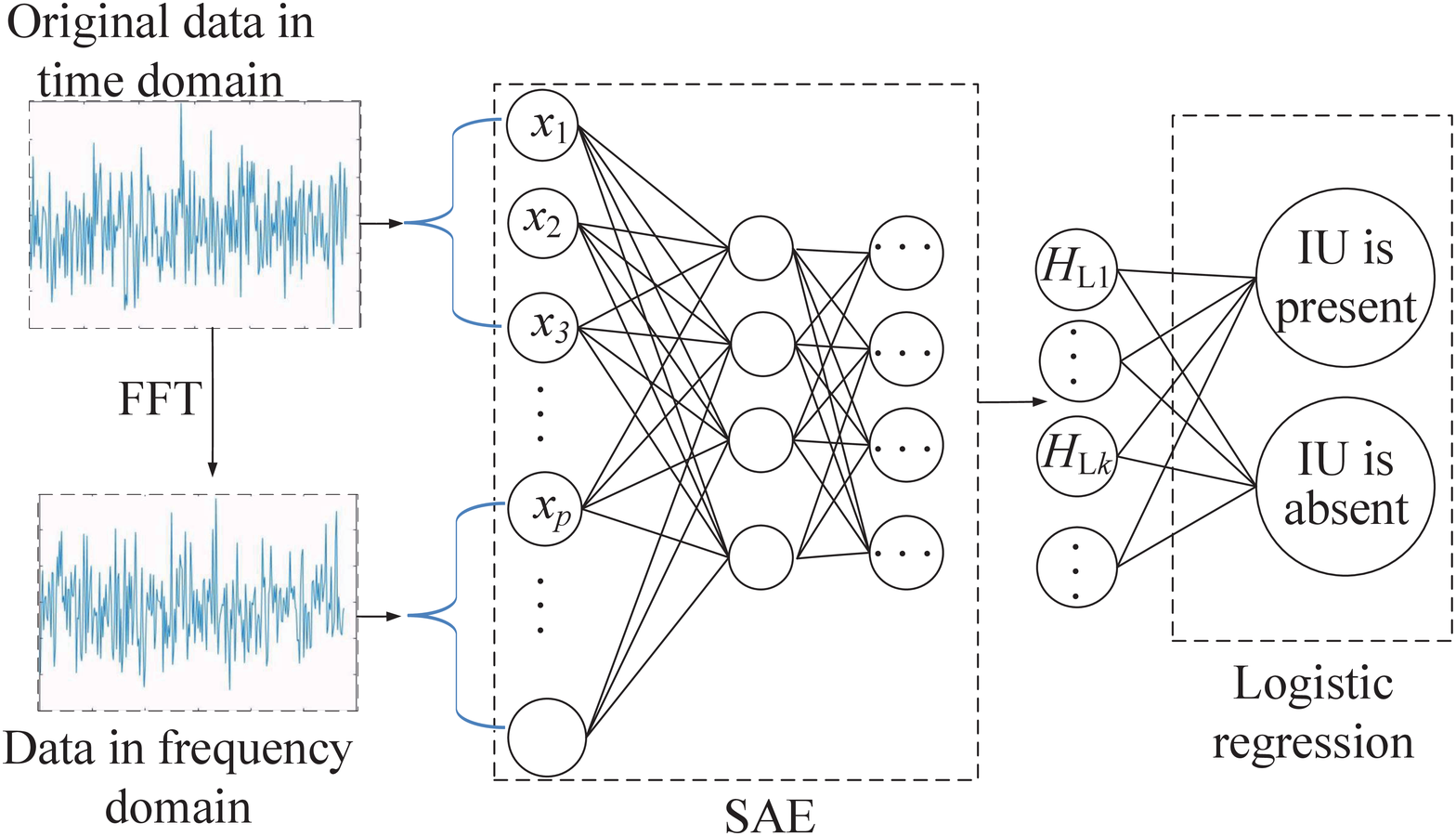}
\DeclareGraphicsExtensions.
\caption{Structure of proposed SAE-TF with logistic regression classifier}
\vspace{-1cm}
\label{SAE-TF}
\end{figure}

The first step of SAE-TF is to transfer the original received signals from the time domain to the frequency domain by Fast Fourier transform (FFT):
\begin{eqnarray}\label{fft_y}
\mathbf{Y}=\mathbf{FFT}(\mathbf{y}),
\end{eqnarray}
where $\mathbf{FFT}(.)$ denotes the FFT operation. Then $\mathbf{Y}$ is divided into real and imaginary parts, respectively. Under this situation, the frequency domain signals are expressed as:
\begin{eqnarray}\label{y_f_i}
\mathbf{Y}_i\!\!\!&:=\!\!&\!\!\!\{\Re({Y}_i(0)),\Im({Y}_i(0)), \dots,
\Re({Y}_i(N_c+N_d-1)),\Im({Y}_i(N_c+N_d-1))\}.
\end{eqnarray}

Thus the frequency domain input vector of SAE-SS network $\mathbf{X}$ can be expressed as:
\begin{eqnarray}\label{input_X}
\mathbf{X}:=\{\mathbf{Y}_0, \mathbf{Y}_1, \dots, \mathbf{Y}_{M-1}\}^T,
\end{eqnarray}
where the length of $\mathbf{X}$ is equal to $2M(N_c+N_d)$.

Then the input signals (the linear arrangement of $\mathbf{x}$ and $\mathbf{X}$), are fed into the SAE network for training, as shown in Fig. \ref{SAE-TF}. After pre-training, fine-tuning of SAE, which are the same as in Section \ref{Proposed Stacked Autoencoder Based Spectrum Sensing Method}, SAE-TF is used to sense IU's presence/absence during the spectrum sensing stage.

Notably, unlike SAE-SS, the input of SAE-TF involves two parts: the original received signal $\mathbf{x}$ and the FFT of received signal $\mathbf{X}$. $\mathbf{x}$ (the only input to SAE-SS) provides the essential information/features in the time domain while $\mathbf{X}$ provides the useful features in the frequency domain. In such a case, SAE-TF can jointly extract more useful features from both time and frequency domains. This makes the two events of interest (IU presence and IU absence) more distinguishable/separate for the classification purpose. SAE-TF is hence able to improve the sensing performance of SAE-SS that only utilizes the time domain features.
\begin{figure*}[tb]
\setlength{\abovecaptionskip}{0.cm}
\setlength{\belowcaptionskip}{-0.cm}
\centering
\subfigure[Output signals of second hidden layer for SAE-SS in two dimensions]{
\begin{minipage}{0.465\textwidth}
\includegraphics[scale=0.5]{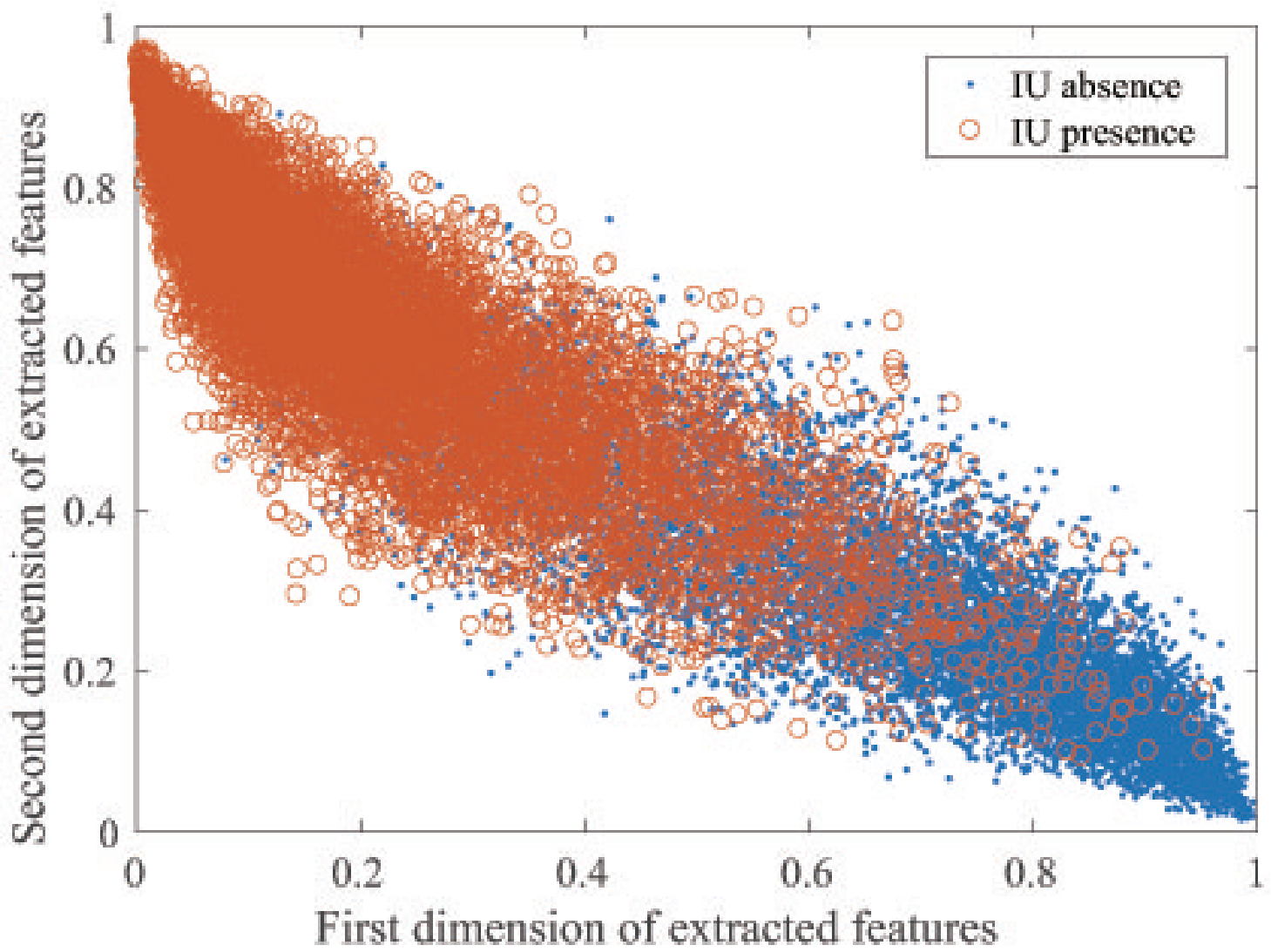}
\label{sae_ss}
\end{minipage}
}
\subfigure[Output signals of second hidden layer for SAE-TF in two dimensions]{
\begin{minipage}{0.465\textwidth}
\includegraphics[scale=0.5]{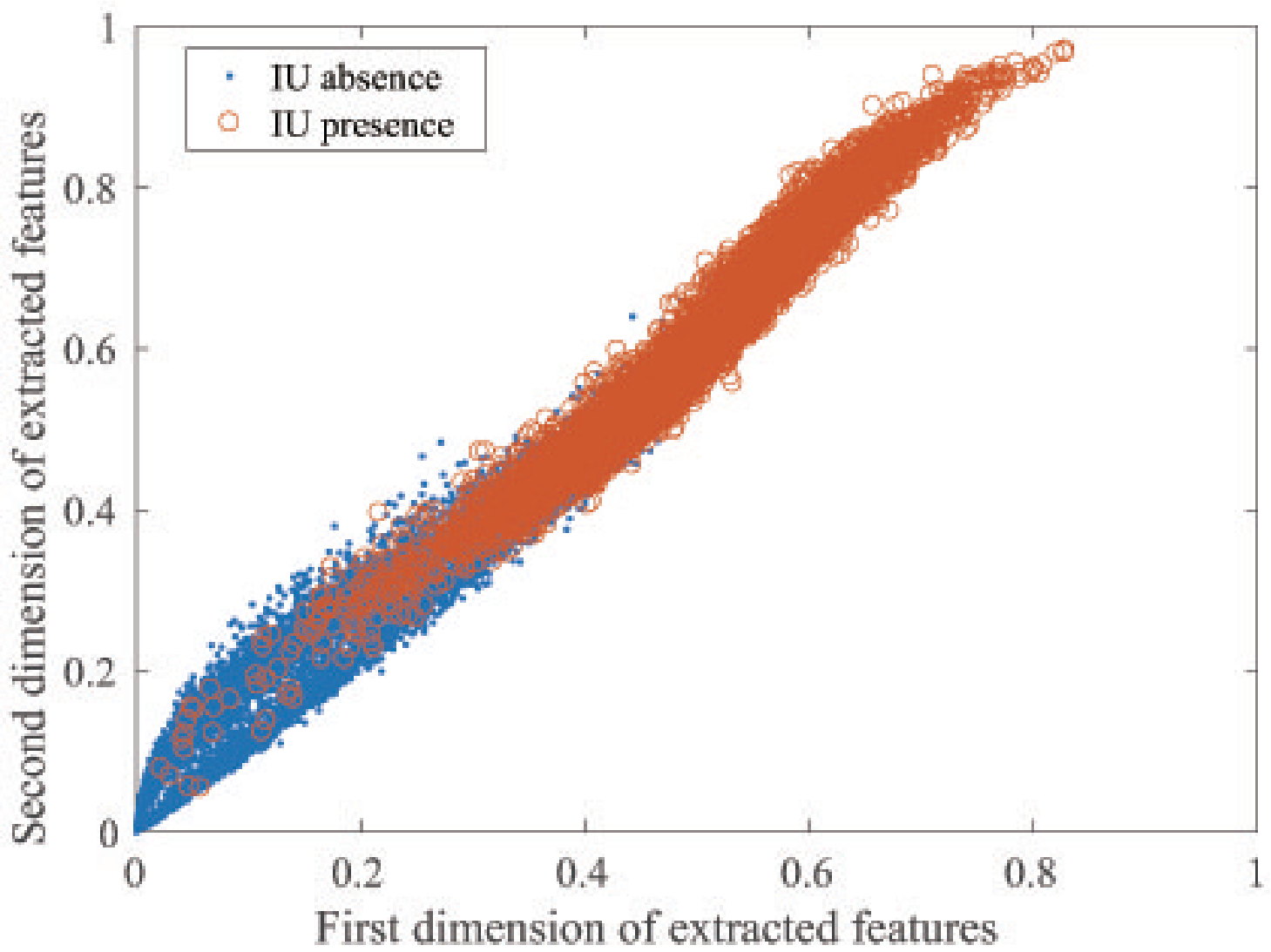}
\label{osae_tf_snr15}
\end{minipage}
}
\caption{Performance of extracting hidden features for SAE-SS and SAE-TF in two dimensions}
\vspace{-1 cm}
\label{hidden features of SAE-SS and SAE-TF}
\end{figure*}

We provide Fig. \ref{hidden features of SAE-SS and SAE-TF} to show the extracted two-dimensional features captured by SAE-SS and SAE-TF. It is clear that for SAE-SS, the distribution of two events ( i.e., representing IU's presence and absence) has larger overlap than that of SAE-TF. In other words, the features of SAE-TF can be easier distinguished into separated classes than those of SAE-SS. Therefore, SAE-TF is able to achieve a better sensing performance than SAE-SS owing to the features extracted from both time and frequency domains. However, the number of input units of SAE-TF doubles that of SAE-SS, so the training complexity is higher (more details in next section and the Section \ref{Simulation Results}). The pseudo-code of SAE-TF is in Algorithm 2.
\begin{table}
\setlength{\abovecaptionskip}{0.cm}
\setlength{\belowcaptionskip}{-0.cm}
\centering
	\label{SAE-TF Steps}
	\vskip 0.3cm
\begin{minipage}[htbp]{3.8 in}
\rule{\linewidth}{0.3mm}\vspace{-0.05in}
{{\bf{Algorithm 2}}: Proposed Stacked Autoencoder Based Spectrum Sensing Method with Time-Frequency Domain Signals (SAE-TF)}\vspace{-0.12in}\\
\rule{\linewidth}{0.2mm}
{ {\small
\begin{tabular}{ll}
   1:&\textbf{begin}\\
   2:& Transfer the received signals $\mathbf{y}$ using FFT and obtain $\mathbf{Y}$\\
   3:& Achieve $\mathbf{x}$ and $\mathbf{X}$ based on $\mathbf{y}$ and $\mathbf{Y}$\\
   4:& \textbf{For} $1\leq l\leq L$\\
   5:& \quad  Build an AE with $N_v$ units of input layer and $N_h$ units\\
     & \quad of hidden layer;  \\
   6:& \quad  \textbf{If} $l=1$\\
   7:& \quad \quad $N_v=2*N_{input}$; $N_h=K_1$;    \\
   8:& \quad \quad  The input signals of the first AE are the linear \\
     & \quad \quad arrangement of $\mathbf{x}$ and $\mathbf{X}$;\\
   9:& \quad \textbf{else} \\
   10:& \quad \quad $N_v=P_l$; $N_h=K_l$;       \\
   11:& \quad \quad the hidden units of previous layer ${\mathbf{H}}_{l-1}$ is set as \\
      & \quad \quad the input of current layer ${\mathbf{H}}_{l}^{in}$;      \\
   12:& \quad \textbf{end} \\
   13:& Train SAE-TF with the input signals. \\
      &Training procedures are the same to Summary 1.\\
   14:&\textbf{end}\\
\end{tabular}}}\\
\rule{\linewidth}{0.3mm}
\end{minipage}\\
\vspace{-0.9cm}
%\caption{Summary of Key Steps in PSEO}
\end{table}

\section{Complexity Analysis and Other Discussions}
\label{Complexity Analysis and Other Discussions}
\subsection{Complexity Analysis}
Table \ref{computation} shows the computational complexity of different sensing methods during the online sensing phase. We take the number of complex multiplication and the real multiplication as metrics because they are the most computationally expensive. Moreover, since one complex multiplication can be treated as four times of real multiplication, we use the total number of real multiplication to show the computational complexity. In Table \ref{computation}, $K_l$ and $K_L$ are the numbers of units in the hidden layer of the $l$th and $L$th layer, respectively. $P_l$ is the number of input units in the $l$th layer. $S_{\delta}$ (denoted as $S_{\tau}$ in \cite{osssoskunv}) is the set of consecutive indices for which $x(n)=x(n+N_d)$, given the synchronization mismatch $\delta$. $N_w$ denotes that the received signal is divided into $N_w$ segments in \cite{8302117}. The size of the convolution kernel is $N_x\times N_y\times N_{ch}$.
\begin{table} [!t]\small
\setlength{\abovecaptionskip}{0.cm}
\setlength{\belowcaptionskip}{-0.cm}
	\centering
	\caption{Online computational complexity of different sensing methods}
	\begin{tabular}{|c|c|c|}
		\hline
		{Method}& Complex Multiplications & Real Multiplication \\
		\hline
		CP                                        & $(N_c+N_d)(M+N_{S_{\tau}}+1)+M(N_c+N_d)^2$ & $2(N_c+N_d)(N_c+N_d-N_{S_{\tau}})$\\
		\hline
		CM                                        &$MN_d\text{log}_2\frac{N_d}{2}+MN_d^2$& $2(N_d^2-N_d)$ \\
		\hline
		ANN                                       &$2M(N_c+N_d)$ & $4K_1+\sum_{l=2}^{L}{K_lP_l}+K_L$\\
		\hline
		CNN                                     &$4M(N_c+N_d)(1+\text{log}_2{(N_d+N_c)})$ & $2N_wN_xN_yN_{ch}+K_L$\\
		\hline
		SAE-SS                                    &None & $2MK_1(N_c+N_d)+\sum_{l=2}^{L}{K_lP_l}+K_L$\\
		\hline
		SAE-TF                                    &$M(N_d+N_c)\text{log}_2{(N_d+N_c)}$ & $4MK_1(N_c+N_d)+\sum_{l=2}^{L}{K_lP_l}+K_L$\\
		\hline
	\end{tabular}
\vspace{-1 cm}
	\label{computation}
\end{table}

From this table, the online computational complexity of SAE-SS is intermediate, and that of SAE-TF is the highest, among these sensing methods. Note that, there is no complex multiplication for SAE-SS because we partition the complex signals into the real and imaginary parts (refer to equation (\ref{y_i})). For SAE-TF, the complex multiplications come from the operation of FFT (refer to equation (\ref{fft_y})). When $N_c=8$, $N_d=64$, $M=2$, $L=2$, $N_{S_\tau}=7$, $K_1=100$, $K_2=50$, $N_w=10$, $N_x=N_y=4$, and $N_{ch}=5$, the total numbers of real complication of the proposed SAE-SS and SAE-TF are $33850$ and $66202$, respectively. By contrast, the numbers of ANN-based, CNN-based, CM-based and CP-based methods are $7754$, $18170$, $43392$, and $53568$, respectively. Besides, the online sensing time of SAE-TF is only about $0.647ms$ in our experiment environment, which is feasible in practice. Moreover, that time can be significantly reduced by using more powerful computing resource in practice. Note that although SAE-TF has the highest number of real multiplication, it achieves much better sensing performance than other methods.
\subsection{SAE-SS and SAE-TF with Additional Input Features}
We further study the impact of additional information/features (e.g., the energy of received signal, CP feature and covariance matrix (CM) feature) on the sensing performance. As an example, for the CP and CM features, we adopt their test statistics from \cite{osssoskunv} and \cite{7265098}, respectively. We take the SAE-SS as a study case (similar results can be easily obtained for SAE-TF). Its new input signal is
\begin{eqnarray}\label{input_x plus AF}
{\mathbf{\hat{x}}}:=\{{\bf{x}}, {\bf{x}}_{af}\},
\end{eqnarray}
where $\bf{x}$ is the initial input signal, and ${\bf{x}}_{af}$ is the additional feature, which can be written as
\begin{eqnarray}\label{additional feature x}
\mathbf{x}_{af}:=\{{\it{\Upsilon}}_{\text{ED}}, {\it{\Upsilon}}_{\text{CP}}, {\it{\Upsilon}}_{\text{CM}}\}^T,
\end{eqnarray}
where ${\it{\Upsilon}}_{\text{ED}}$, ${\it{\Upsilon}}_{\text{CP}}$ and ${\it{\Upsilon}}_{\text{CM}}$ are test statistics of ED, CP- and CM-based detections, respectively.

As observed in the Fig. \ref{PM AF} in Section \ref{Simulation Results}, additional feature to the input can help SAE-SS and SAE-TF reduce the PM. This implies that considering the additional features can improve the sensing performance of our proposed methods. However, that is achieved at the expense of higher complexity (in both the offline training and online spectrum sensing stages).
\subsection{Training and other Practical Aspects}
Like other neural network-based methods (e.g., ANN-based \cite{8292449} and CNN-based \cite{8302117}), SAE-SS and SAE-TF require previous data for the training stage. At the beginning when there is no previous monitoring data, to collect and record the monitoring data for the later training purpose, SUs can use the conventional sensing methods to detect IU's presence. After sufficient monitoring data are recorded, SUs can proceed with our proposed methods (by first using these data to train the SAE). In the next section, we also numerically study different training methods (by adapting or not adapting with the received signal SNR levels).

Note that though SAE-SS and SAE-TF require training data, they don't require any prior knowledge about the IUs' signal. That makes SAE-SS and SAE-TF very much practical, especially for sharing spectrum bands used in military (e.g., 3500-3650 MHz band in SAS \cite{SAS_LSA}). It is also apparent that SAE-SS and SAE-TF require higher computational complexity than conventional OFDM signal sensing methods (e.g., basic ED method, CP-based sensing method \cite{osssoskunv}, CM-based sensing method \cite{7265098}). However, rapid advances in specialized hardware for ML-based computation (e.g., GPU circuits) would allow radio devices to accommodate SAE-SS and SAE-TF. Moreover, the computation time involved in the offline training stage (required just once) is well-amortized in many subsequent online IU sensing instances.

\section{Simulation Results and Comparison}
\label{Simulation Results}
In this section, we first conduct simulations to analyze the performance of the proposed two methods (e.g., SAE-SS and SAE-TF) and compare them with conventional OFDM signal sensing methods (e.g., basic ED method, CP-based sensing method \cite{osssoskunv}, CM-based sensing method \cite{7265098}) and neural network-based methods (ANN-based \cite{8292449},  {{and CNN-based methods \cite{8302117}}}). Then we provide the performance analysis of SAE-SS and SAE-TF with different parameters. The propagation channel between IU transmitter and SU receiver is a frequency selective Rayleigh fading channel. For both the training phase and sensing phase, we generate an OFDM system with the binary phase shift keying (BPSK) modulation \cite{6910238}. The OFDM block size of IU signals is $N_d=64$ and CP length is $N_c=N_d/8$. The signal bandwidth is $5$ MHz, and the radio frequency is $2.4$ GHz. Each subcarrier spacing is $78.125$ kHz, and the duration for each OFDM sample is $12.8$ $\mu s$. In both the training phase and the online sensing phase, SNR varies from $-20$dB to $-8$dB. The timing delay is $\delta \in [0, N_c+N_d-1]$. The normalized CFO is $f_d \in [0,1]$. The SNR, timing delay and CFO in the training phase are the same as the sensing phase.

We set the number of hidden layers for both SAE-SS and SAE-TF is $L=2$. The first and the second hidden layer contain $100$ and $50$ units, respectively. The number of received OFDM blocks is $M=2$. The number of input units for SAE-SS and SAE-TF are $2M(N_c+N_d)$ and $2*2M(N_c+N_d)$, respectively. The probability of false alarm is PFA$= 0.05$. Both the training data set and the testing data set contain $2*10^4$ samples. The number of iterations is $5000$. We use TensorFlow 1.3 with Python language to train the proposed methods. The experiments are conducted on the $6$-Core $3.6$GHz PC with Nvidia P$5000$ graphic card (16GB memory).

Fig. \ref{performance_of SAE-SS and SAE-TF} shows the performance of extracted hidden/unknown features for SAE-SS and SAE-TF under SNR$=-15$dB, compared with the input signals of the first layer. Specifically, the first $10^4$ samples denote ``IU is absent" and the second $10^4$ samples mean ``IU is present". The numbers of input data of the first layer for SAE-SS and SAE-TF are $288$ and $576$, respectively. For each method, the number of output signals of the second hidden layer is $50$. From Fig. \ref{input_first_layer_new_pluse_snr15} and Fig. \ref{input_first_layer_new_pluse_fft_snr15}, it is very difficult to differentiate between ``IU is absent" and ``IU is present" based on the input signals only. However, they can be readily separated by the output signals of the second hidden layer as (much more clearly) visually observed in Fig. \ref{output_last_layer_new_snr15} and Fig. \ref{output_last_layer_new_pluse_fft_snr15}. We interpret that is because SAE-SS and SAE-TF extract more hidden features from the received signals.
\begin{figure*}[tb]
\setlength{\abovecaptionskip}{0.cm}
\setlength{\belowcaptionskip}{-0.cm}
\centering
\subfigure[]{
\begin{minipage}{0.225\textwidth}
\includegraphics[scale=0.32]{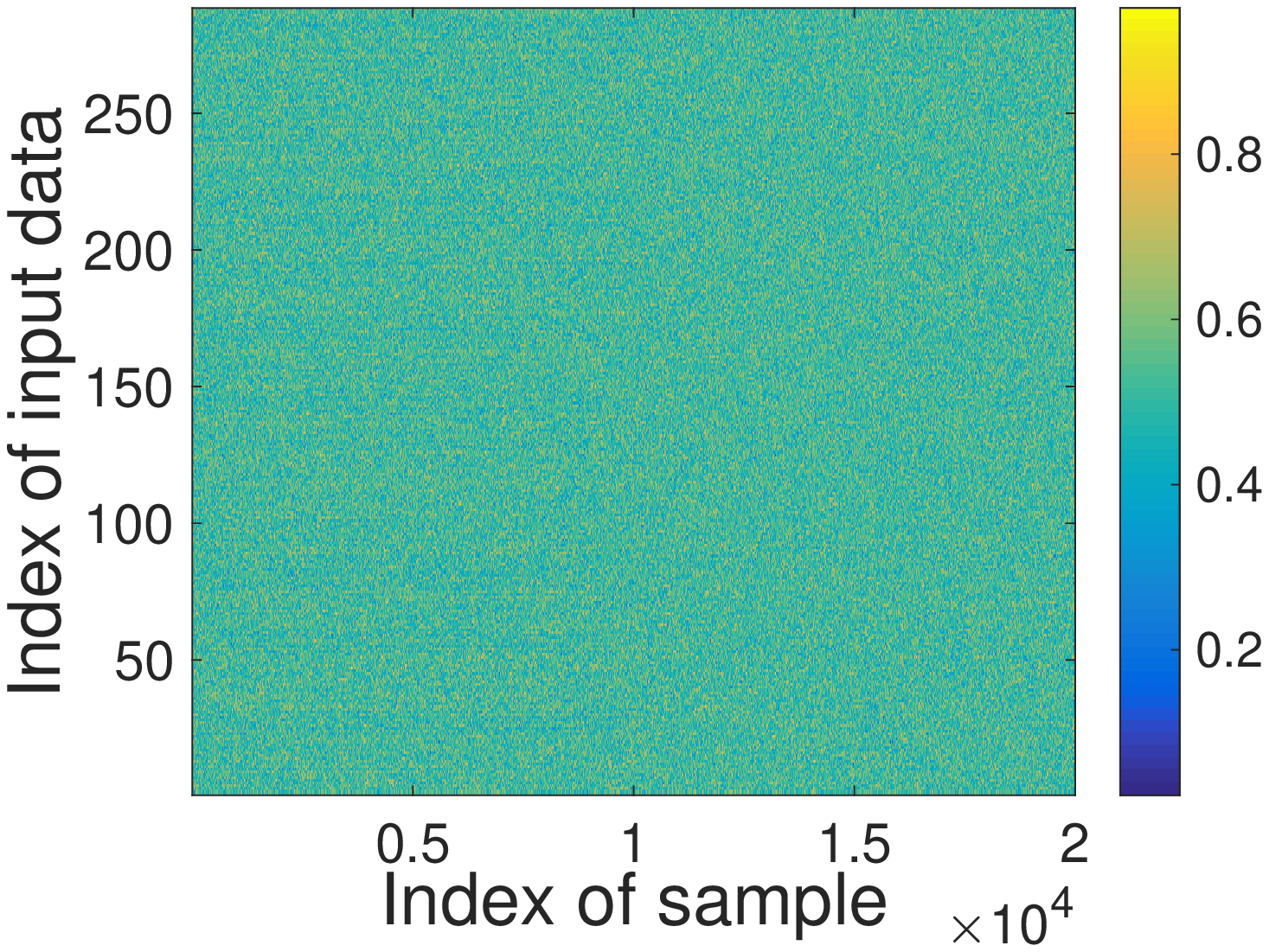}
\label{input_first_layer_new_pluse_snr15}
\end{minipage}
}
\subfigure[]{
\begin{minipage}{0.225\textwidth}
\includegraphics[scale=0.32]{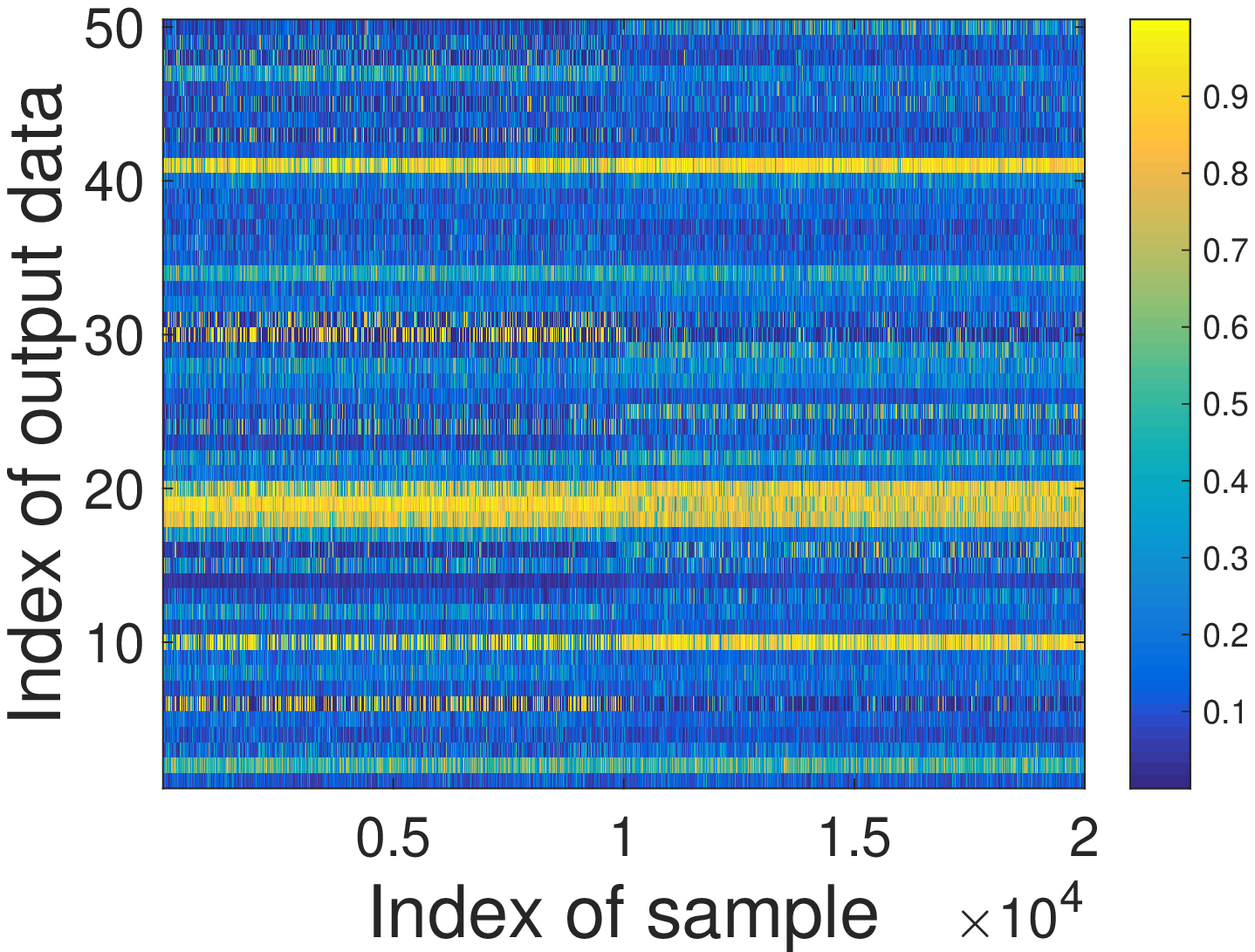}
\label{output_last_layer_new_snr15}
\end{minipage}
}
\subfigure[]{
\begin{minipage}{0.225\textwidth}
\includegraphics[scale=0.32]{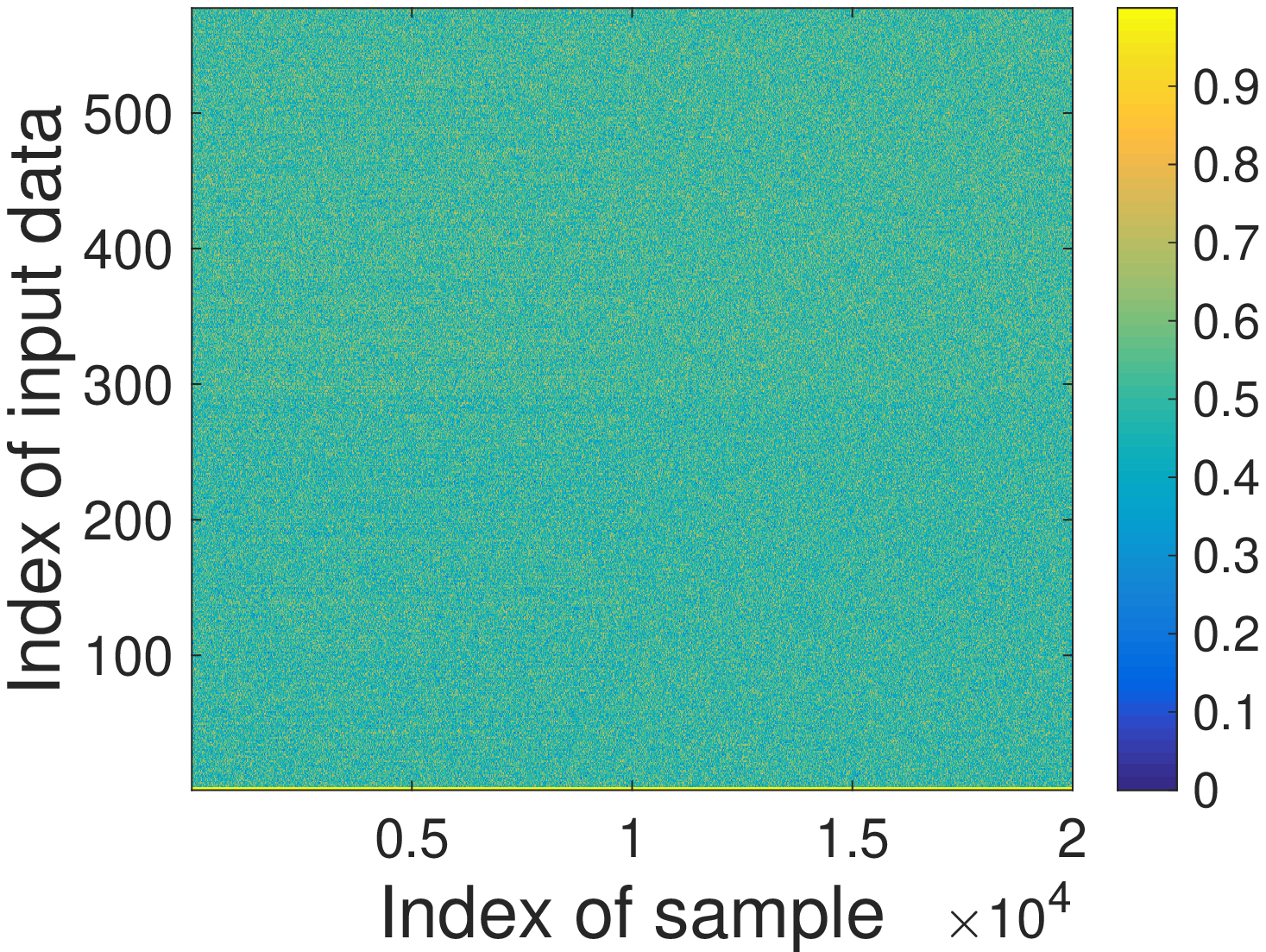}
\label{input_first_layer_new_pluse_fft_snr15}
\end{minipage}
}
\subfigure[]{
\begin{minipage}{0.225\textwidth}
\includegraphics[scale=0.32]{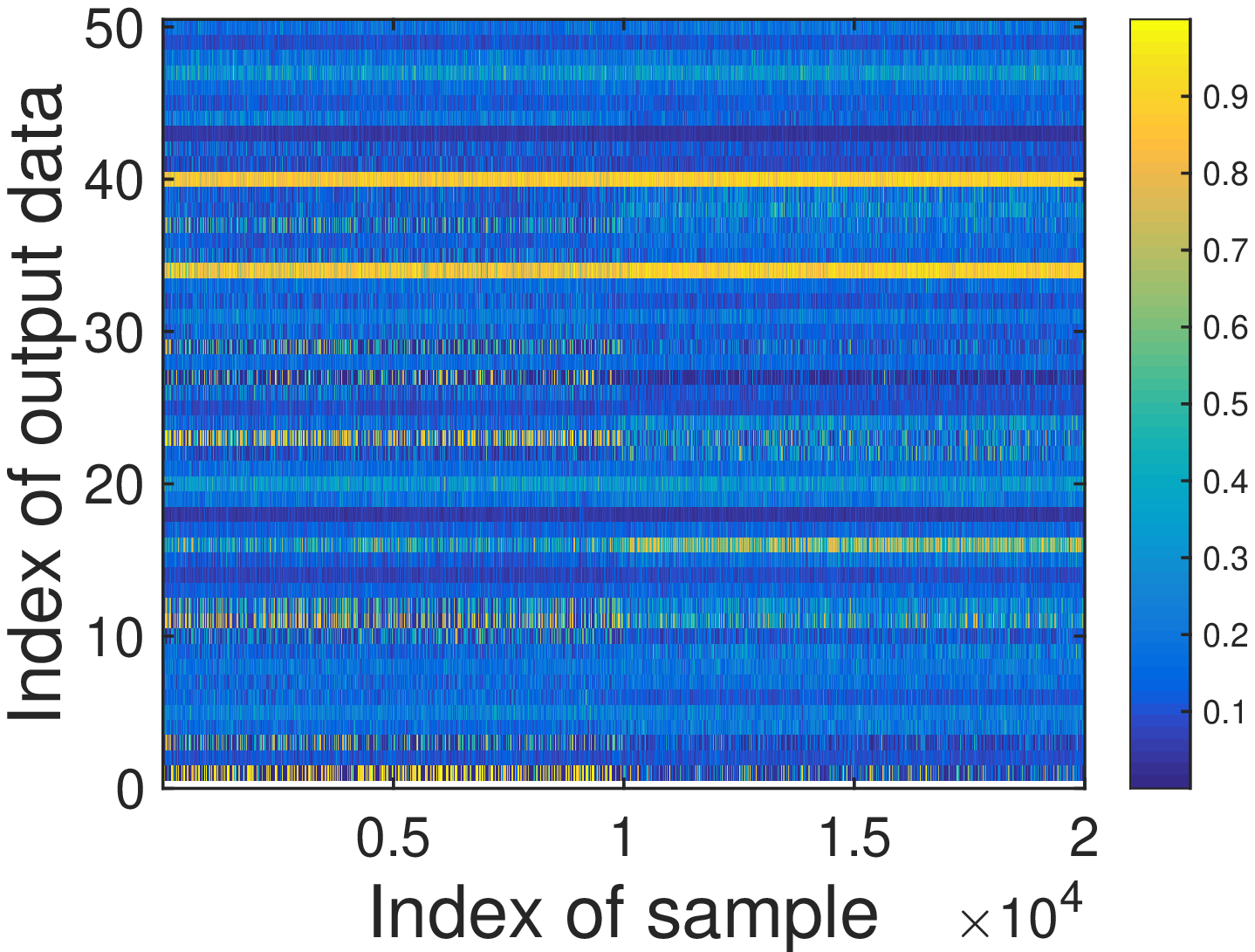}
\label{output_last_layer_new_pluse_fft_snr15}
\end{minipage}
}
\caption{The performance of extracting hidden features for SAE-SS and SAE-TF under SNR=-15dB. (a) Input signals of the first layer for SAE-SS; (b) Output signals of the second hidden layer for SAE-SS; (c) Input signals of the first layer for SAE-TF; (d) Output signals of the second hidden layer for SAE-TF.}
\vspace{-1cm}
\label{performance_of SAE-SS and SAE-TF}
\end{figure*}

%\begin{figure*}[tb]
%\centering
%\subfigure[Input signals of the first layer for SAE-SS]{
%\begin{minipage}{0.22\textwidth}
%\includegraphics[scale=0.21]{input_first_layer_new_pluse_snr15.eps}
%\label{input_first_layer_new_pluse_snr15}
%\end{minipage}
%}
%\subfigure[Output signals of the second hidden layer for SAE-SS]{
%\begin{minipage}{0.22\textwidth}
%\includegraphics[scale=0.21]{output_last_layer_new_snr15.eps}
%\label{output_last_layer_new_snr15}
%\end{minipage}
%}
%\subfigure[Input signals of the first layer for SAE-TF]{
%\begin{minipage}{0.22\textwidth}
%\includegraphics[scale=0.21]{input_first_layer_new_pluse_fft_snr15.eps}
%\label{input_first_layer_new_pluse_fft_snr15}
%\end{minipage}
%}
%\subfigure[Output signals of the second hidden layer for SAE-TF]{
%\begin{minipage}{0.22\textwidth}
%\includegraphics[scale=0.21]{output_last_layer_new_pluse_fft_snr15.eps}
%\label{output_last_layer_new_pluse_fft_snr15}
%\end{minipage}
%}
%\caption{The performance of extracting hidden features for SAE-SS and SAE-TF under SNR=-15dB}
%\label{performance_of SAE-SS and SAE-TF}
%\end{figure*}

\subsection{Training Strategies for SAE-SS and SAE-TF}
For our proposed schemes, we can use one of the two training strategies to train the SAE network. For the training strategy $1$ (TS-$1$), we divide the training data into different groups with respect to the SNR conditions to train the different SAE architectures \cite{5601105,8302117}. The architecture parameters are held unchanged after the training process. During the spectrum sensing stage, depending on the level of the SNR, different SAE architectures can be selected. This adaptation can significantly improve the sensing performance.

For the training strategy $2$ (TS-$2$), we use all the training data to train a single SAE network for all different SNR conditions. It means the trained SAE is later used without requiring SNR values. However, the sensing performance of TS-$2$ is worse than that of TS-$1$, under the same training conditions, as shown in Fig. \ref{orignal_with_different_ts_lb}.
\begin{figure}[tb]
\setlength{\abovecaptionskip}{0.cm}
\setlength{\belowcaptionskip}{-0.cm}
\centering
\includegraphics[width=8 cm]{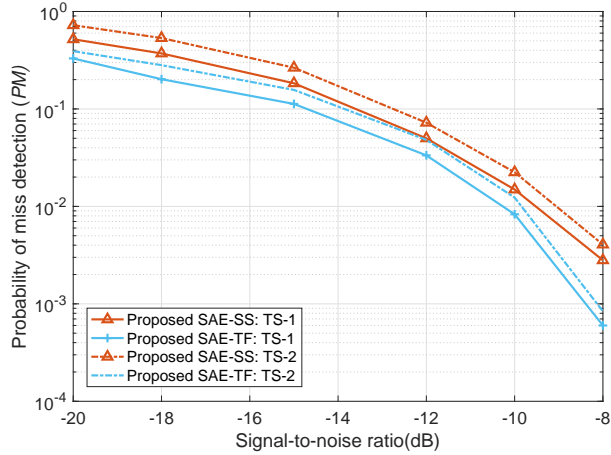}
\DeclareGraphicsExtensions.
\caption{Probability of missed detection for SAE-SS and SAE-TF using two training strategies under perfect condition}
\vspace{-1cm}
\label{orignal_with_different_ts_lb}
\end{figure}

Fig. \ref{orignal_with_different_ts_lb} compares the sensing performance of TS-$1$ and TS-$2$ under the ``perfect condition". In the perfect condition, there is no noise uncertainty, timing delay or carrier frequency offset. Additionally, the SU has sufficient prior knowledge of IU signals (e.g., signal structure, CP of IU's signals, transmitting power, and noise power). In this figure, it is clear that our proposed SAE-SS and SAE-TF can achieve the smaller PM using the strategy TS-$1$. Therefore, in the sequel, we use the strategy TS-$1$ to train SAE networks. It is important to note that though TS-$1$ can achieve better sensing performance than TS-$2$, the accurate noise power is required for selecting the trained SAE architectures. Thus the noise uncertainty affects the performance of TS-$1$ (refer to Fig. \ref{Noise_uncertainty_new}). By contrast, TS-$2$ is immune to the noise uncertainty, as it does not need the noise power for sensing purpose.

% One can also apply the data augment technique \cite{Krizhevsky2012ImageNet,Salamon2016Deep} to generate additional synthetic training data from the previous monitoring data, enlarging the size of the training dataset. We then put the enlarged training dataset into the SAE network to learn and extract their essential features.

\subsection{Comparison with Existing Conventional and Neural Network based OFDM Sensing Methods}
In this subsection, we compare sensing performance and training time under different conditions for seven sensing methods.
\begin{figure}[tb]
\setlength{\abovecaptionskip}{0.cm}
\setlength{\belowcaptionskip}{-0.cm}
\centering
\includegraphics[width=8 cm]{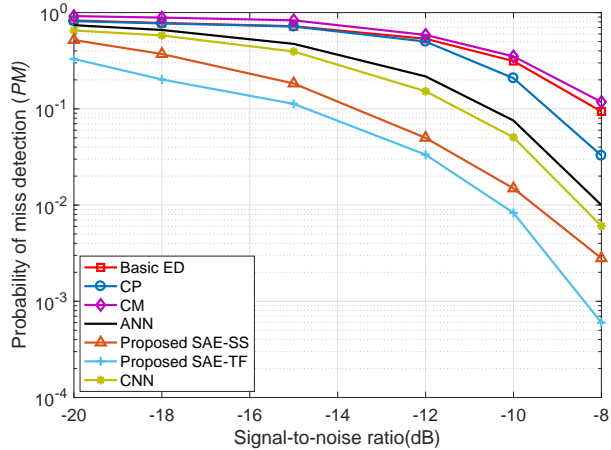}
\DeclareGraphicsExtensions.
\caption{Probability of missed detection among different spectrum sensing methods under perfect conditions}
\vspace{-1cm}
\label{orignal_new}
\end{figure}
\subsubsection{Sensing performance} Under ``perfect condition", as can be seen in Fig. \ref{orignal_new}, the values of PM of SAE-SS and SAE-TF are much smaller than all other sensing methods. For instance, when SNR=$-20$dB, the PM of SAE-SS and SAE-TF are only $0.5189$ and $0.3298$, respectively. By contrast, the figures for ED, CM-based, CP-based, ANN-based, and CNN-based sensing methods are $0.8279$, $0.9192$, $0.8145$, $0.7413$, and $0.652$, respectively. Moreover, with the increase of SNR, the PM of SAE-SS and SAE-TF reduce much faster than the other sensing methods.
\begin{figure}[t]
\setlength{\abovecaptionskip}{0.cm}
\setlength{\belowcaptionskip}{-0.cm}
\centering
\includegraphics[width=8 cm]{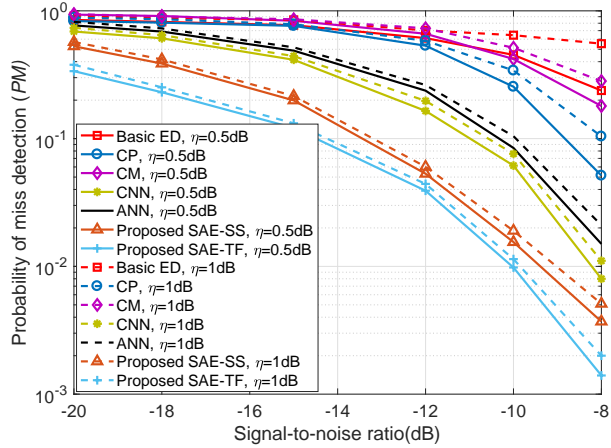}
\DeclareGraphicsExtensions.
\caption{Probability of missed detection among different spectrum sensing methods with noise uncertainty}
\vspace{-1 cm}
\label{Noise_uncertainty_new}
\end{figure}

Fig. \ref{Noise_uncertainty_new} shows the impact of noise uncertainty $\eta$ on seven sensing methods. According to this figure, it is clear that the proposed SAE-SS and SAE-TF are much more robust to noise uncertainty than the other five sensing methods. For instance, when $\eta$ increases from $0.5$dB to $1$dB under SNR$=-10$dB, PM of SAE-SS and SAE-TF only increase from $0.0155$ to $0.0190$ and from $0.0098$ to $0.0114$, respectively. On the contrary, PM of CNN-based, ANN-based, CP-based, CM-based, and ED methods increase from $0.0615$ to $0.0761$, from $0.0845$ to $0.1045$, from $0.2549$ to $0.3415$, from $0.4213$ to $0.5142$, and from $0.4522$ to $0.6450$, respectively.
\begin{figure}[tb]
\setlength{\abovecaptionskip}{0.cm}
\setlength{\belowcaptionskip}{-0.cm}
\centering
\includegraphics[width=8 cm]{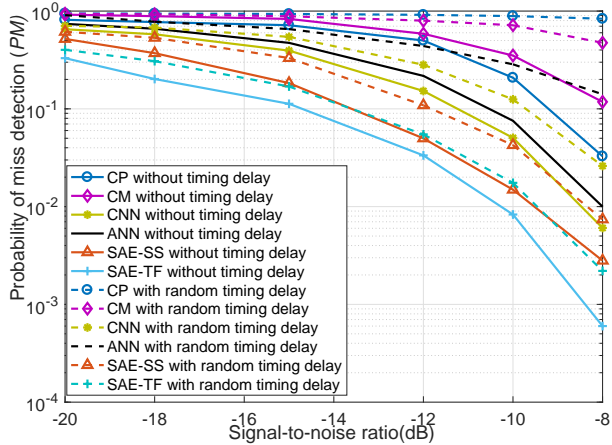}
\DeclareGraphicsExtensions.
\caption{Probability of missed detection for SAE-SS and SAE-TF with random timing delay}
\vspace{-0.8cm}
\label{time_delay_rand}
\end{figure}

Fig. \ref{time_delay_rand} compares the effect of random timing delay on the sensing performance of different methods. Since ED is not affected by timing delay, we only present the sensing performance of the other six sensing methods. In this figure, the timing delay is uniformly distributed in the range $[0, N_c+N_d-1]$, where $N_d=64$, $N_c=8$. As can be seen from this figure, compared with the existing methods, our proposed SAE-SS and SAE-TF are more robust to the timing delay. Moreover, our proposed methods are able to achieve a much smaller PM than the other methods. For instance, when SNR=$-12$dB, the PM of SAE-SS and SAE-TF are $0.1088$ and $0.0548$, respectively. However, the PM of CNN-based, ANN-based, CM-based and CP-based are $0.2832$, $0.4393$, $0.798$ and $0.9155$, respectively.
\begin{figure}[tb]
\setlength{\abovecaptionskip}{0.cm}
\setlength{\belowcaptionskip}{-0.cm}
\centering
\includegraphics[width=8 cm]{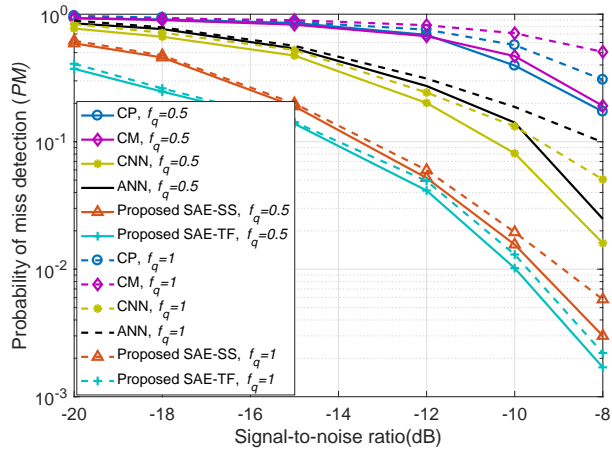}
\DeclareGraphicsExtensions.
\caption{Probability of missed detection among different spectrum sensing methods with CFO}
\vspace{-1cm}
\label{cfo_new}
\end{figure}

Fig. \ref{cfo_new} shows the impact of CFO. The normalized CFO in this figure is set to $0.5$ and $1$, respectively.  The sensing results of ED are not presented in this figure, because they are not affected by CFO. It is obvious that the proposed SAE-SS and SAE-TF outperform the CP-based, CM-based, ANN-based, and CNN-based sensing methods regarding the robustness to CFO. SAE-SS and SAE-TF are capable of achieving much smaller PM than the other four sensing methods. Moreover, the PM of SAE-SS and SAE-TF increase slightly when the normalized CFO grows from $0.5$ to $1$. However, PM of the ANN-based, CM-based and CP-based methods increase significantly.

As can be seen from Fig. \ref{orignal_new} $\sim$ Fig. \ref{cfo_new}, our SAE-based methods outperform ANN-based \cite{8292449} and CNN-based \cite{8302117} sensing methods. This is thanks to the fact that the input data in our proposed methods are the original received data, containing much more essential information about IU's activity. By contrast, the ANN-based method only uses the energy and Zhang test statistic \cite{Zhang2005Likelihood} from Likelihood Ratio Test statistic as features to train the ANN network. For CNN-based method, it only uses the energy and cyclostationary features to train the CNN for detecting IUs' presence. As such, both ANN- and CNN-based methods inevitably lose other important (but hidden) features of IU's activity (in addition to what were used).

\subsubsection{Training time}
\begin{table*}\small
\setlength{\abovecaptionskip}{0.cm}
\setlength{\belowcaptionskip}{-0.cm}
  \centering
  \caption{Performance and training time of different learning-based methods}
  \begin{tabular}{|*{8}{c|}}
    \hline
        \multirow{2}*{Method }  & \multicolumn{3}{|c|}{Offline Training Time} &  \multicolumn{3}{|c|}{PM (SNR=$-12$dB)}  \\
         \cline{2-7}                  &$L_{TS}=10^3$  & $L_{TS}=10^4$ & $L_{TS}=10^5$ &$L_{TS}=10^3$ & $L_{TS}=10^4$ & $L_{TS}=10^5$  \\ \hline
        SAE-SS                        & 0.32mins      & 2.87mins      & 24.61mins       & 0.081    & 0.0499        & 0.0445        \\ \hline
        SAE-TF                        & 0.59mins      & 5.81mins      & 43.68mins       & 0.072    & 0.0334        & 0.0264     \\ \hline
        ANN                           & 0.12mins      & 0.97mins      & 8.21mins        & 0.524    & 0.2713        & 0.2132      \\ \hline
        CNN                           & 0.21mins      & 1.72mins      & 15.74mins       & 0.388    & 0.1473        & 0.1121 \\ \hline
  \end{tabular}
  \vspace{-1cm}
\label{performance and training time}
\end{table*}
Table \ref{performance and training time} shows the relationship between the sensing performance and the training time with different amount of training samples. In this table, $L_{TS}$ stands for the number of OFDM samples in the training phase. As can be seen from this table, the more training samples, the lower PM (i.e., the higher sensing accuracy). Although our proposed sensing methods require a relatively large amount of data for training, the training phase is conducted offline, i.e., before the spectrum sensing stage. Hence the training phase does not occupy the time or resource of the spectrum sensing stage. Moreover, the offline training is only required once. After that, SUs can use the trained network for sensing IU's activity in the future as long as user's operational information doesn't change. As an example, in this paper, the signal bandwidth is $5$MHz, and the radio frequency is $2.4$GHz. Each subcarrier spacing is $78.125$ kHz, and the duration for each OFDM sample is $12.8\mu$s. Thus it is feasible to obtain the required amount of samples for our proposed approaches. Besides, the user can select different amount of samples based on its specific requirement. Notably, that training time can be significantly reduced with the specialized hardware implementation/design in practice (e.g., using specialized GPU cards from NVIDIA).% One can also use further reduce the number of hidden units and the number of hidden layers, to speed up the decision process of the proposed methods..
\subsection{Sensing Performance of SAE-SS and SAE-TF under Different Settings and Additional Features}
In this subsection, we further analyze the sensing performance of the proposed SAE-SS and SAE-TF under different settings: hidden layers, hidden units, received OFDM blocks, CP lengths, and with additional features.
\begin{figure}[tb]
\setlength{\abovecaptionskip}{0.cm}
\setlength{\belowcaptionskip}{-0.cm}
\centering
\includegraphics[width=8 cm]{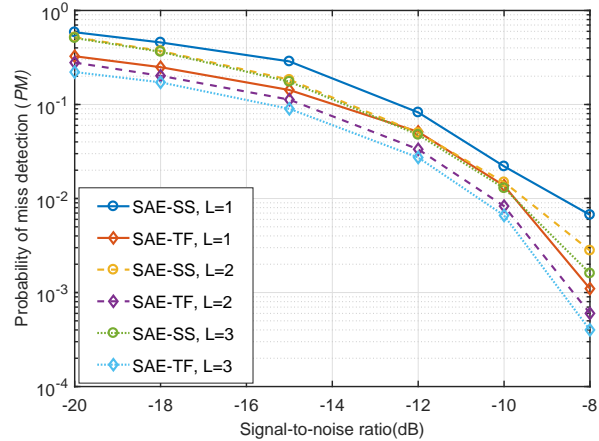}
\DeclareGraphicsExtensions.
\caption{Probability of missed detection of SAE-SS and SAE-TF with different number of hidden layers}
\vspace{-1cm}
\label{diff_hidden_layer}
\end{figure}

Fig. \ref{diff_hidden_layer} shows the impact of the number of hidden layers on SAE-SS and SAE-TF. The number of hidden layers is selected as $L=1,2,3$, and the corresponding number of units in hidden layers are $(100)$, $(100,50)$ and $(100, 50, 20)$, respectively. The number of OFDM blocks is $M=2$. According to this figure, the values of PM for these two sensing methods are smaller by increasing $L$. For example, when $L$ increases from $1$ to $3$, SNR=$-20$dB, the PM of SAE-SS and SAE-TF decrease from $0.5871$ to $0.4731$ and from $0.3260$ to $0.2213$, respectively. However, the complexities of these two methods would also increase, meaning that a specific $L$ should be selected based on different circumstances.
\begin{figure}[tb]
\setlength{\abovecaptionskip}{0.cm}
\setlength{\belowcaptionskip}{-0.cm}
\centering
\includegraphics[width=8 cm]{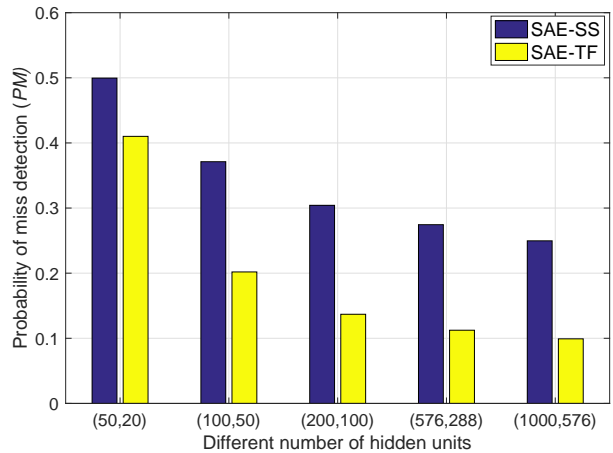}
\DeclareGraphicsExtensions.
\caption{Probability of missed detection for SAE-SS and SAE-TF with different number of hidden units}
\vspace{-1.1cm}
\label{diff_hidden_units_new}
\end{figure}

Fig. \ref{diff_hidden_units_new} shows the effect of different number of hidden units on the sensing performance. In this figure, SNR$=-18$dB, $M=2$ and $L=2$, the numbers of input units for SAE-SS and SAE-TF are $288$ and $576$, respectively. When the number of hidden units is equal or larger than that of input units, PM is slightly reduced, meaning that the sensing accuracy is better. However, accordingly, the SAE network also becomes much more complicated, significantly increasing the complexity of the offline training and online sensing stages. Therefore, users can select different number of hidden units based on their specific requirement of sensing accuracy and complexity.
\begin{figure}[tb]
\setlength{\abovecaptionskip}{0.cm}
\setlength{\belowcaptionskip}{-0.cm}
\centering
\includegraphics[width=8 cm]{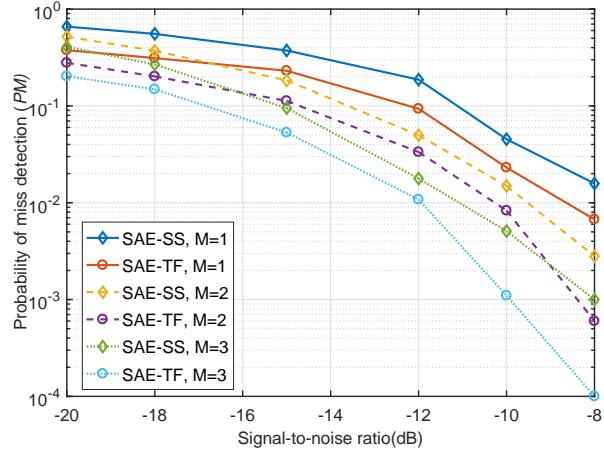}
\DeclareGraphicsExtensions.
\caption{Probability of missed detection of SAE-SS and SAE-TF with different numbers of the received OFDM blocks}
\vspace{-1cm}
\label{diff_M}
\end{figure}
\begin{table} [tb]\small
\vspace{0.2cm}
\setlength{\abovecaptionskip}{0.cm}
\setlength{\belowcaptionskip}{-0.cm}
\centering
\caption{Offline Training time for SAE-SS and SAE-TF}
\begin{tabular}{|c|c|c|c|}
\hline
{\backslashbox{ Methods }{M}} & $M=1$  & $M=2$ & $M=3$
\\
\hline
SAE-SS  & $104.68$mins  & $159.32$mins  &$254.15$mins \\
\hline
SAE-TF        & $122.47$mins   & $321.92$mins & $451.24$mins \\
\hline
\end{tabular}
\vspace{-0.8cm}
\label{running time}
\end{table}
Fig. \ref{diff_M} shows the sensing performance of SAE-SS and SAE-TF with different numbers of the received OFDM blocks $M$. According to this figure, it is obvious that the sensing performance of these two sensing methods is affected by the factor $M$. When $M$ changes from $1$ to $3$, SNR=$-10$dB, the PM of SAE-SS declines by $0.0402$, and the PM of SAE-TF decreases by $0.0221$. Notably, the complexity also increases with the growth of $M$, captured by the training time recorded in Table \ref{running time}.

Table \ref{running time} shows the offline training time of SAE-SS and SAE-TF with a different number of OFDM blocks $M$. The number of hidden layers is $L=2$, and there are $100$ and $50$ units in the first and second hidden layer, respectively. The training data set contains $10^6$ samples. The number of iterations is $20000$. Based on this table, the offline training time of SAE-SS is bigger than that of SAE-SS. Moreover, with the increase of $M$, the offline training time of SAE-TF is increasingly longer than SAE-SS. Since SAE-TF can achieve higher sensing performance than SAE-SS, it provides a better tradeoff between accuracy and complexity.

Table \ref{different CP} shows the effect of different CP lengths ($N_c$) on the sensing performance, with SNR$=-12$dB. From this table, it is clear that for both SAE-SS and SAE-TF, the longer CP allows us to achieve better sensing accuracy (i.e., the lower PM value). This implies that our proposed methods do extract more essential information from CP content. For instance, when $N_c$ increases from $0$ to $16$, the PM of SAE-SS and SAE-TF decrease from $0.066$ to $0.0301$ and from $0.0498$ to $0.0211$, respectively. However, it is important to note that the longer CP would lead to a smaller size of data block, reducing the efficiency of data transmission.
\begin{table} [!t]\small
\setlength{\abovecaptionskip}{0.cm}
\setlength{\belowcaptionskip}{-0.cm}
\centering
\caption{The impact of different CP lengths ($N_c$) on the sensing performance}
\begin{tabular}{|c|c|c|c|}
\hline
\diagbox[width=3.7cm]{Method}{CP length}&$N_c=0$&$N_c=8$&$N_c=16$      \\
\hline
SAE-SS                                    &0.066 & 0.0499&  0.0301\\
\hline
SAE-TF                                    &0.0498 & 0.0334& 0.0211 \\
\hline
\end{tabular}
\vspace{-1cm}
\label{different CP}
\end{table}
\begin{figure}[tb]
\setlength{\abovecaptionskip}{0.cm}
\setlength{\belowcaptionskip}{-0.cm}
	\centering
	\includegraphics[width=8 cm]{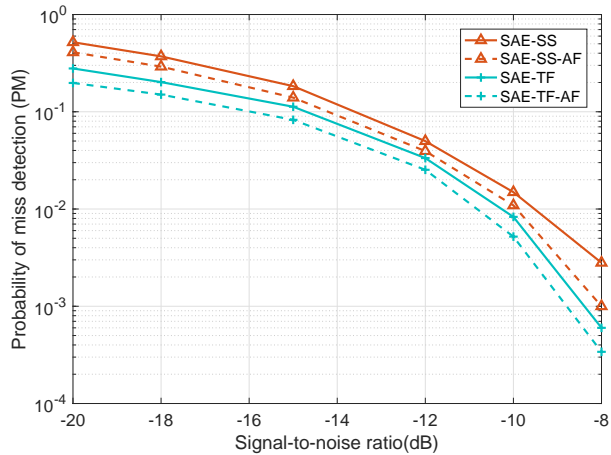}
	\DeclareGraphicsExtensions.
	\caption{Probability of miss detection for SAE-SS and SAE-TF with additional features under perfect condition}
\vspace{-1cm}
	\label{PM AF}
\end{figure}
Fig. \ref{PM AF} depicts the PM of our proposed methods with additional features. As can be seen from this figure, the values of PM of SAE-SS with additional features (SAE-SS-AF) and SAE-TF with additional features (SAE-TF-AF) are smaller than those of SAE-SS and SAE-TF. This implies that considering the additional features can improve the sensing performance of our proposed methods. However, adding additional features would lead to much higher computational complexity for the online sensing phase due to the feature extraction process (refer to Table \ref{computation} for more details). Therefore, users can select different input signal taking their requirements of sensing performance and computational complexity into account.
\section{Conclusion}
\label{Conclusion}
In this paper, we proposed a Stacked Autoencoder Based Spectrum Sensing Method (SAE-SS) and a Stacked Autoencoder Based Spectrum Sensing Method with time-frequency domain signals (SAE-TF) to detect the activity states of IUs using OFDM signal. SAE-SS and SAE-TF are more robust to timing delay, CFO, and noise uncertainty, compared with the conventional OFDM sensing methods. Moreover, they are able to detect IU's activities solely based on the received signals and without any requirement of prior knowledge of IU's signals. SAE-SS and SAE-TF also do not require any external feature extraction algorithms. SAE-TF achieves a better sensing accuracy than SAE-SS, especially under low SNR conditions, while it has the higher complexity. Extensive simulation results demonstrate that SAE-SS and SAE-TF are capable of achieving much higher sensing performance than traditional OFDM sensing methods even under low SNR and severe timing delay, CFO, and noise uncertainty conditions. This is thanks to the capability of the underlying deep neural networks of SAE-SS and SAE-TF that extract both known and unknown hidden features of OFDM signals.

% or


\appendices

%\section*{Acknowledgment}

%\ifCLASSOPTIONcaptionsoff
%  \newpage
%\fi



\bibliographystyle{IEEEtranTCOM}
\bibliography{IEEEabrv,reference}

%\begin{biography}[{\includegraphics[width=1in,height=1.25in,clip,keepaspectratio]{mshell}}]{Michael Shell}


%\vfill


\end{document}